\documentclass[runningheads]{elsarticle}
\usepackage{graphicx}
\usepackage{comment}
\usepackage{envmath}
\usepackage{amsfonts}
\usepackage{amsmath}
\usepackage{amssymb}
\usepackage{multirow}
\usepackage{adjustbox}

\usepackage{xcolor}

\newcommand{\ff}[1]{{#1}}
\newcommand{\etg}[1]{{#1}}
\newcommand{\ovr}[1]{{#1}}

\def\rsp{r_{sp}}
\def\rsc{r_{sc}}
\def\Cah{C_{ah}}
\def\Ksc{K_{sc}}
\def\Ksp{K_{sp}}
\def\Cab{C_{ab}}
\def\Ah{[Amp_h]}
\def\sc0{[S_c]^0}
\def\sp0{[S_p]^0}
\def\C1{C_1}
\def\kc{k_c}

\def\url#1{\texttt{#1}}

\newcommand{\sout}[1]{{}}

\journal{Theoretical Computer Science C natural computing}

\begin{document}

\begin{frontmatter}

\title{A Skin Microbiome Model \ff{with AMP interactions and \sout{with Mathematical}} Analysis of Quasi-Stability vs Stability in Population Dynamics}

\author[l1,l2]{Eléa Thibault Greugny}%\inst{1,}\inst{2} %\orcidID{0000-0001-5812-3722} 
\author[l2]{François Fages}%\inst{2} %\orcidID{0000-0001-5650-8266}
\author[l3]{Ovidiu Radulescu}
\author[l4]{Peter Szmolyan}
\author[l1]{Georgios N. Stamatas}%\inst{1} %\orcidID{0000-0003-4544-7597}

%
%\authorrunning{E. Thibault Greugny et al.}

\address[l1]{Johnson \& Johnson Santé Beauté France, Issy-les-Moulineaux, France}
\address[l2]{Inria Saclay, Lifeware Team, Palaiseau, France}
\address[l3]{University of Montpellier, France}
\address[l4]{Technical University, Wien, Austria}

%
%\maketitle              % typeset the header of the contribution
%
\begin{abstract}
The skin microbiome plays an important role in the maintenance of a healthy skin. It is an ecosystem, composed of several species, competing for resources and interacting with the skin cells. Imbalance in the cutaneous microbiome, also called dysbiosis, has been correlated with several skin conditions, including acne and atopic dermatitis. Generally, dysbiosis is linked to colonization of the skin by a population of opportunistic pathogenic bacteria. Treatments consisting in non-specific elimination of cutaneous microflora have shown conflicting results. 
\sout{It is therefore necessary to understand the factors influencing shifts of the skin microbiome composition.} 
In this article, we introduce a mathematical model based on ordinary differential equations, with 2 types of bacteria populations (skin commensals and opportunistic pathogens) \ff{and including the production of antimicrobial peptides} to study the mechanisms driving the dominance of one population over the other. 
By using published experimental data, assumed to correspond to the observation of stable states in our model, we reduce 
the number of parameters of the model from 13 to 5. \ff{We then use a formal specification in quantitative temporal logic to calibrate our model by global parameter optimization and perform sensitivity analyses}.
On the time scale of 2 days of the experiments,
\ff{the model predicts} that certain changes of the environment, like the elevation of skin surface pH, create favorable conditions for the emergence and colonization of the skin by the opportunistic pathogen population\ff{, while the production of human AMPs has non-linear effect on the balance between pathogens and commensals}.
\ff{\sout{Such predictions help identifying potential therapeutic targets for the treatment of skin conditions involving dysbiosis of the microbiome.}}
\ff{Surprisingly}, simulations on longer time scales reveal that the \ff{equilibrium} reached around 2 days 
%following the introduction of bacteria in the model, 
\ff{can in fact be} a quasi-stable state followed by the reaching of a reversed stable state after 12 days \ff{or more}.
We analyse the conditions of quasi-stability observed in this model using tropical algebraic methods, and
show their non-generic character in contrast to slow-fast systems.
\ff{These conditions are then generalized} to a large class of  population dynamics models over any number of species.
\end{abstract}

\begin{keyword}
skin microbiome;
 atopic dermatitis;
 antimicrobial peptides;
 population dynamics;
 quasi-stability;
 meta-stability;
 ODE models;
 model reduction;
 steady-state reasoning;
 \ff{quantitative temporal logics;}
 \ff{sensitivity analyses;}
 tropical algebra.
\end{keyword}

\end{frontmatter}
%
%
%

%\ff{in addition to direct modifications invitation to contribute with comments using macros ff etg ovr and ps}

\section{Introduction}
Located at the interface between the organism and the surrounding environment, the skin constitutes the first line of defense against external threats, including irritants and pathogens. In order to control potential colonization of the skin surface by pathogens, the epidermal cells, called keratinocytes, produce antimicrobial peptides (AMPs)~\cite{pazgier_human_2006}. The physiologically acidic skin surface pH also contributes to control the growth of bacterial populations~\cite{proksch_ph_2018,korting_differences_1990}. Another contributor to the defense against pathogen colonization are commensal bacteria in the community of microorganisms living on the skin, commonly referred to as the skin microbiome. Over the past decade, several studies have highlighted the key role played by such commensal bacterial species defending against invading pathogens, as well as their contribution to the regulation of the immune system~\cite{lai_commensal_2009,cogen_staphylococcus_2010,lai_activation_2010,kong_skin_2011,belkaid_dialogue_2014,byrd_human_2018}.

Alterations in the composition of the skin microbiome resulting in a dominance by a pathogenic species, also called dysbiosis, have been associated with skin conditions such as acne or atopic dermatitis (AD)~\cite{leyden_propionibacterium_1975,kong_temporal_2012}. In the case of AD, the patient skin is often colonized by \textit{Staphylococcus aureus} (\textit{S. aureus}), especially on the lesions~\cite{kong_temporal_2012},
while in the case of acne it is colonized by another bacteria, \textit{C. acnes}. Treatment strategies targeting non-specific elimination of cutaneous microflora, such as bleach baths, have shown conflicting results regarding their capacity to reduce the disease severity~\cite{chopra_efficacy_2017}. On the other hand, treatments involving introduction of commensal species, like \textit{Staphylococcus hominis}~\cite{nakatsuji_development_2021} on the skin surface appear promising. Accordingly, the interactions between the commensal populations, pathogens and skin cells seem at the heart of maintaining microbiome balance. There is therefore a necessity to investigate further those interactions and the drivers of dominance of one population over others. Unfortunately, it is challenging to perform \textit{in vitro} experiments involving more than one or two different species, even more so on skin explants or skin equivalents. 

Mathematical models of population dynamics have been developed and used for more that 200 years~\cite{malthus_essay_1798}. 
Here\footnote{This article is an extended version of a previous communication at CMSB 2022 published in~\cite{GSF22cmsb}. The \ff{equilibrium\sout{stabilization}} constraint\ff{, formalized in quantitative temporal logic and} used here for parameter search and \ff{sensitivity} analyses, is a stronger property than in the previous version. \ff{Results of global parameter optimization and sensitivity analyses have now} been added for all parameters. The main addition however is the mathematical analysis of the quasi-stability phenomenon exhibited by the model, the analysis of its non-generic character, in contrast to well-studied slow-fast systems, and \ff{its} generalization
  %of the presented conditions on the parameter values for exhibiting such quasi-stability behaviours
  to a large class of population dynamics models over any number of species in Sec.~\ref{sec:math}.\sout{ is a new section devoted to that mathematical analysis and generalization.}}, we introduce a model based on ordinary differential equations (ODEs), describing the interactions of a population of commensal species with one of opportunistic pathogens and the skin cells. \etg{To our knowledge, there are only two other mathematical models describing the interactions between different bacteria populations and the skin~\cite{nakaoka2016, miyano2022}}. Nakaoka et al. introduced a model combining ODE and delayed differential equations to look at the competition dynamic between 2 populations of bacteria exposed to cytokines. Miyano et al. designed a QSP model including the interactions between \textit{S. aureus} and coagulase negative Staphylococcus to evaluate AD treatment strategies targeting specifically \textit{S. aureus}. \ff{To the best of our knowledge, the model we present here} is the first to consider explicitly the production of antimicrobial peptides by skin commensal bacteria. This \ff{allows us} to study the factors influencing the dominance of one population over the other\etg{, \ff{especially the elevation of} pH \ff{and the production of} antimicrobial peptides, \ff{in addition to the respective growth rates of the bacterial populations.}} 
\sout{on a microbiologically relevant timescale of a couple of days
corresponding to biological experimental data.}

More specifically, \sout{we identify constraining relationships on the parameter values,} based on published experimental data of Nakatsuji et al.
\cite{nakatsuji_antimicrobials_2017} and Kohda et al.~\cite{kohda_vitro_2021}, 
\ff{assumed to correspond to steady states} of our \ff{parametric ODE} model,
we \ff{derive constraints \sout{identify constraining relationships}} on the parameter values
which allow us to reduce 
the parametric dimension of our model from 13 to 5 parameters.
\ff{The landscape of those 5 remaining parameter values is then explored using global parameter optimization and sensitivity analyses procedures,
  implemented in our modeling software Biocham~\cite{CFS06bi},
  using a specification of the properties of the different bacterial population equilibria} formalized in quantitative temporal logic FO-LTL($\mathbb{R}_{lin}$)~\cite{RBFS11tcs}.

\ff{On} the time scale of the experiments of Kohda et al.~\cite{kohda_vitro_2021},
\ff{our model predicts} that certain changes in the environment, like an elevation of skin surface pH, create favorable conditions for the emergence and colonization of the skin by the opportunistic pathogen population.
\ff{The model also predicts that the concentration of human AMPs affects in a non-linear manner the balance between pathogens and commensals.}
Such predictions \ff{can} help identify potential therapeutic strategies for the treatment of skin conditions involving microbiome dysbiosis\ff{, or optimize AMP levels on the skin surface in topical treatments}.

Interestingly, in the reduced model, we observe an unexpected phenomenon of
meta-stability~\cite{bovier1970metastability,TK08neuron,RSNGW15cmsb,samal2016geometric,samal2018metastable,desoeuvres_CMSB2022} or quasi-stability~\cite{morozov2020long}, 
a terminology we prefer to adopt in the absence of stochasticity,
in which the seemingly stable state \ff{observed at} the time scale of the experiments of 2 days
is followed by a switch after 12 days toward a reversed stable state.
We develop a complete mathematical analysis of the conditions of quasi-stability observed in our population dynamics model using singular perturbations and tropical algebraic methods.
We show the non-generic character of this form of quasi-stability in our model, in sharp contrast to slow-fast systems which are mainly studied.
Finally, we generalize the conditions \ff{first} obtained \ff{by tropical analysis of} the parameter values of our model, to a large class of population dynamics models over any number of species which will similarly exhibit quasi-stability phenomena on relatively long timescales.

\section{Initial ODE model with 13 parameters}

\begin{figure}
    \centering
    \includegraphics[width=0.7\textwidth]{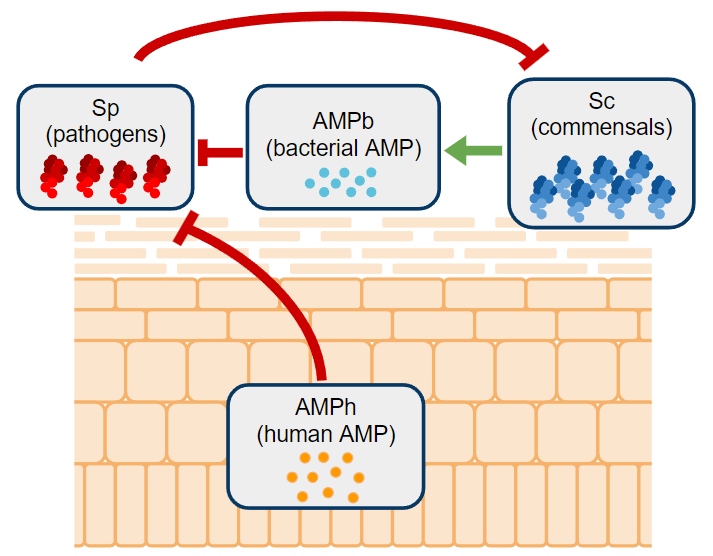}
    \caption{Model overview \ff{of the interactions between the commensal and pathogenic bacterial populations at the surface of the skin, including the production of bacterial and human AMPs.} Green arrows \ff{represent positive production effects} and red T-lines represent \ff{negative} killing effect.}
    \label{fig:ModelOverview}
\end{figure}

The \ff{ODE} model built in this article considers two types of bacterial populations. The first population, $S_c$, regroups commensal bacteria species having an overall beneficial effect for the skin, and the second population, $S_p$, represents opportunistic pathogens. \ff{One originality of our model is that it considers their interactions through the production of both human and bacterial AMPs, as depicted in Fig.~\ref{fig:ModelOverview}.}

The \ff{terms of the } differential equations \ff{governing the growth of } both bacterial populations are based on the common logistic growth model~\cite{zwietering_modeling_1990}, considering non-explicitly the limitations in food and space. The limited resources are included in the parameters $K_{sc}$ and $K_{sp}$, representing the optimum concentration of the populations in a given environment, considering the available resources. \etg{$r_{sc}$ and $r_{sp}$ represent the growth rate of commensal and pathogenic bacteria respectively.}

The bactericidal effect of antimicrobial peptides (AMPs) produced by skin cells, $Amp_h$, on $S_p$ is included with a Hill function \ff{of order 1 representing a simple saturation effect. Parameter} \etg{$d_{sph}$ represents the maximal killing rate of $S_p$ by $Amp_h$ and $C_{ah}$ is the concentration of $Amp_h$ inducing half this maximum killing rate.} This type of highly non-linear functions \etg{is commonly\sout{have been}} used \etg{\sout{previously}} to model the \etg{saturating} effect of antibiotics on bacterial populations \etg{\cite{campion_pharmacodynamic_2005,meredith_bacterial_2015,spalding2018}}. For the sake of simplicity, the AMPs produced by skin cells is introduced as a constant parameter, $[Amp_h]$, in the model. It represents the average concentration of these AMPs among surface cells, under given human genetic background and environmental conditions.

Several studies revealed that commensal bacterial populations, like \textit{S. epidermidis} or \textit{S. hominis}, are also able to produce AMPs targeted against opportunistic pathogens, such as \textit{S. aureus}~\cite{cogen_selective_2010,nakatsuji_antimicrobials_2017}.
For these reasons, we \ff{chose to also} introduce in \ff{our} model AMPs of bacterial origin, $Amp_b$, acting similarly to $Amp_h$ on the pathogenic population $S_p$. $Amp_b$ is produced at rate $k_c$ by $S_c$, and degraded at rate $d_a$. Furthermore, we include a defense mechanism of $S_p$ against $S_c$ with a direct killing effect. \etg{This killing effect is also modelled with a Hill function \etg{or order 1}, where $d_{sc}$ is the maximum killing rate of $S_c$ by $S_p$ and $C_1$ the concentration of $S_p$ necessary to reach half this maximum killing rate.} 

Altogether, this gives us the following ODE system with 3 variables and 13 parameters, all taking non-negative values:

\begin{System}\label{fullSystem}
\frac{d [S_c]}{dt} = \left( r_{sc} \left( 1 - \frac{[S_c]}{K_{sc}} \right) - \frac{d_{sc} [S_p]}{C_1 + [S_p]} \right) [S_c] \\ \\

\frac{d [S_p]}{dt} = \left(r_{sp} \left( 1 - \frac{[S_p]}{K_{sp}} \right) - \frac{d_{spb} [Amp_b]}{C_{ab} + [Amp_b]} - \frac{d_{sph} [Amp_h]}{C_{ah} + [Amp_h]} \right) [S_p]\\ \\

\frac{d [Amp_b]}{dt} = k_c [S_c] - d_a [Amp_b]
\end{System}
\sout{The model is illustrated on Fig.~\ref{fig:ModelOverview} and }Table \ref{tab13param} recapitulates the \ff{3} variables and \ff{13} parameters of this model with their units.

\begin{table}
\caption{List of the parameters and variables of our mathematical model with their units. CFU = Colony forming unit, AU = Arbitrary Unit, ASU = Arbitrary Surface Unit}\label{tab13param}

\begin{tabular}[H]{|c|p{10cm}|}
\hline
\textbf{Variable} &  \textbf{Interpretation (unit)}\\
\hline
$[S_c]$ & Surface apparent concentration of $S_c$ ($CFU.ASU^{-1}$)\\
$[S_p]$ & Surface apparent concentration of $S_p$ ($CFU.ASU^{-1}$)\\
$[Amp_b]$ & Concentration of $Amp_b$ ($AU.ASU^{-1}$)\\
\hline
\textbf{Parameter} & \textbf{Interpretation (unit)}\\
\hline
$r_{sc}$ & Growth rate of $S_c$ ($h^{-1}$) \\
$r_{sp}$ & Growth rate of $S_p$, ($h^{-1}$) \\
$K_{sc}$ & Optimum concentration of $S_c$ ($CFU.ASU^{-1}$) \\
$K_{sp}$ & Optimum concentration of $S_p$ ($CFU.ASU^{-1}$) \\
$d_{sc}$ & Maximal killing rate of $S_c$ by $S_p$ ($h^{-1}$) \\
$C_1$ & Concentration of $S_p$ inducing half the maximum killing rate $d_{sc}$ ($CFU.ASU^{-1}$) \\
$d_{spb}$ & Maximal killing rate of $S_p$ by $Amp_b$, ($h^{-1}$) \\
$C_{ab}$ & Concentration of $Amp_b$ inducing half the maximum killing rate $d_{spb}$ ($AU.ASU^{-1}$) \\
$d_{sph}$ & Maximal killing rate of $S_p$ by $Amp_h$, ($h^{-1}$) \\
$C_{ah}$ & Concentration of $Amp_h$ inducing half the maximum killing rate $d_{sph}$ ($AU.ASU^{-1}$) \\
$[Amp_h]$ & Concentration of AMPs produced by the skin cells ($AU.ASU^{-1}$)\\
$k_c$ & Production rate of $Amp_b$ by $S_c$ ($AU.h^{-1}.CFU^{-1}$) \\
$d_a$ & Degradation rate of $Amp_b$ ($AU.h^{-1}$)\\
\hline
\end{tabular}
\end{table}

\sout{Such a model cannot be solved analytically. Furthermore, the use of optimization algorithms to infer the 13 parameter values from data resulted in many valid sets of parameter values. Therefore, it is clearly necessary to restrict the number of parameters by identifying some of them, to be able to analyze the model.}

\section{Using published experimental data to \ff{eliminate \sout{define relations between}} model parameters by steady-state reasoning}
The amount of quantitative experimental data available for the model calibration is very limited due to the difficulty of carrying out experiments involving co-cultures of different bacterial species. Most of the published work focuses on single species or on measuring the relative abundances of species living on the skin, which is highly variable between individuals and skin sites~\cite{grice_topographical_2009}. In the case of AD specifically, \textit{S. aureus} is considered pathogenic and \textit{S. epidermidis} commensal. Published data exist however for those species which we can use to constrain the parameter values of the model.

Two series of \textit{in vitro} experiments are considered~\cite{nakatsuji_antimicrobials_2017,kohda_vitro_2021}. While \textit{in vitro} cultures, even on epidermal equivalent, do not entirely capture the native growth of bacteria on human skin, they provide useful quantitative data that would be very difficult to measure \textit{in vivo}.

In the first experiment~\cite{kohda_vitro_2021}, mono-cultures and co-cultures of \textit{S. epidermidis} and \textit{S. aureus} were allowed to develop on a 3D epidermal equivalent. 
Table \ref{tabExpDataKohda} recapitulates the population sizes of the two species measured after 48 hours of incubation. Kohda \textit{et al.}~also performed another co-culture experiment where \textit{S.epidermidis} was inoculated 4 hours prior to \textit{S.aureus} in the media. This data is not used here as it requires additional manipulation to match the situation represented by the model. However, it would be interesting to use it in the future for model validation.

In the second experiment~\cite{nakatsuji_antimicrobials_2017} the impact of human (LL-37) and bacterial (\textit{Sh}-lantibiotics) AMPs on \textit{S. aureus} survival was studied. The experiments were performed \textit{in vitro}, and the \textit{S. aureus} population size was measured after 24 hours of incubation.
Table \ref{tabExpDataNakatsuji} summarizes their observations.

\begin{table}
\caption{Experimental data from Kohda et. al~\cite{kohda_vitro_2021} used for identifying parameter values. 
}\label{tabExpDataKohda}
\begin{center}
\begin{tabular}[H]{l|c|c}
& \textbf{\textit{S. epidermidis} (CFU/well)} & \textbf{\textit{S. aureus} (CFU/well)}\\
\hline
\textbf{Mono-cultures} & $4.10^8$ & $3.10^9$ \\
\textbf{Co-cultures} & $1.10^8$ & $1.10^9$ \\
\end{tabular}
\end{center}
\end{table}

\begin{table}
\caption{Experimental data from Nakatsuji et. al~\cite{nakatsuji_antimicrobials_2017} used for identifying parameter relations. 
}\label{tabExpDataNakatsuji}
\begin{center}
\begin{tabular}[H]{c|c|c}
 \textbf{\textit{Sh}-lantibiotics ($\mu M$)} & \textbf{LL-37 ($\mu M$) }& \textbf{\textit{S. aureus} (CFU/mL)}\\
\hline
0 & 4 & $10^9$\\
0 & 8 & $6.10^5$\\
0.32 & 0 & $5.10^8$\\
0.64 & 0 & $3.10^3$
\end{tabular}
\end{center}
\end{table}

\subsection{Parameter values inferred from mono-culture experiment data} \label{subsec:Param_mono}
We consider first the monocultures experiments from Kohda \textit{et al.}~\cite{kohda_vitro_2021}, representing the simplest experimental conditions. \textit{S. epidermis} is a representative of the commensal population $S_c$, and \textit{S. aureus} of the pathogenic one, $S_p$. 
Since the two species are not interacting, the set of equations simplifies to:

\begin{System}
\frac{d [S_c]}{dt} = \left( r_{sc} \left( 1 - \frac{[S_c]}{K_{sc}} \right) \right) [S_c] \\ \\

\frac{d [S_p]}{dt} = \left(r_{sp} \left( 1 - \frac{[S_p]}{K_{sp}} \right) \right) [S_p]
\end{System}

At steady-state, the population concentrations are either zero, or equal to their optimum capacities ($K_{sc}$ or $K_{sp}$) when the initial population concentration is non-zero. Given the rapid growth of bacterial population, the experimental measurements done after 48 hours of incubation can be considered as corresponding to a steady-state, which gives:

\begin{equation} \label{Kse}
K_{sc} = 4.10^8 \; CFU.ASU^{-1}
\end{equation}
\begin{equation} \label{Ksa}
K_{sp} = 3.10^9 \; CFU.ASU^{-1}
\end{equation}

\subsection{Parameter relations inferred from experimental data on AMP}
The experimental conditions of Nakatsuji \textit{et al.}~\cite{nakatsuji_antimicrobials_2017} correspond to the special case where there is no commensal bacteria alive in the environment, only the bacterial AMPs, in addition to those produced by the skin cells. Our system of equations then reduces to:

\begin{equation}
\frac{d [S_p]}{dt} = \left(r_{sp} \left( 1 - \dfrac{[S_p]}{K_{sp}} \right) - \frac{d_{spb} [Amp_b]}{C_{ab} + [Amp_b]} - \frac{d_{sph} [Amp_h]}{C_{ah} + [Amp_h]} \right) [S_p]
\end{equation}

The concentrations in LL-37 and \textit{Sh}-lantibiotics, translated in our model into $[Amp_h]$ and $[Amp_b]$ respectively, are part of the experimental settings. Therefore, we consider them as constants over time. At steady state, we get:

\begin{equation} \label{NakaEqSS}
[S_p]^* = 0 \quad \textrm{or} \quad [S_p]^* = K_{sp} \left ( 1 - \frac{d_{spb} [Amp_b]}{r_{sp} (C_{ab} + [Amp_b])} - \frac{d_{sph} [Amp_h]}{r_{sp} (C_{ah} + [Amp_h])} \right)
\end{equation}

Let us first focus on the special case where no \textit{Sh}-lantibiotics were introduced in the media, translating into $[Amp_b] = 0$ in our model. We consider again that the biological observations after 24 hours of incubation correspond to steady-state and substitute the experimental values measured \\
$[Amp_h] = 4 \,\mu M$ ; $[S_p]^* = 10^9$ CFU, and $[Amp_h] = 8 \,\mu M$ ; $[S_p]^* = 6.10^5$ CFU, \\
together with the values of $K_{sc}$ and $K_{sp}$ (from \eqref{Kse} and \eqref{Ksa}) in \eqref{NakaEqSS}, to obtain the following equations:

\begin{System}
\frac{d_{sph}}{r_{sp}} = \frac{4 +C_{ah}}{6} \\ \\
\frac{d_{sph}}{r_{sp}} = \frac{(10^4 - 2)(C_{ah} +8)}{8.10^4}
\end{System}
which reduce to $C_{ah} = 8$ and $\displaystyle{\frac{d_{sph}}{r_{sp}} = 2}$.\\

Following the same method with the experimental conditions without any LL-37 (i.e. $[Amp_h] = 0$) and using two data points \\
($[Amp_b] = 0.32 \,\mu M$ ; $[S_p]^* = 5.10^8$ CFU)\\ and ($[Amp_b] = 0.64 \; \mu M$ ; $[S_p]^* = 3.10^3$ CFU), \\
we get $C_{ab} = 0.16$ and $\displaystyle{\frac{d_{spb}}{r_{sp}} = \frac{5}{4}}$.

It is notable that the maximum killing rates of $S_p$ by $Amp_b$ and $Amp_h$ are both proportional to $S_p$ growth rate. Interestingly, such proportional relation has been observed experimentally between the killing rate of \textit{Escherichia coli} by an antibiotic and the bacterial growth rate~\cite{tuomanen_rate_1986}.

To be consistent with the ranges of \textit{Sh}-lantibiotics concentrations described in Nakatsuji \textit{et al.}~\cite{nakatsuji_antimicrobials_2017}, $[Amp_b]$ should take positive values below 10. Given that $[Amp_b]^* =\displaystyle{ \frac{k_c [S_c]^*}{d_a}}$ at steady-state, and that $K_{sc} = 4.10^8$ CFU is the upper bound for $[S_c]^*$, we obtain the following constraint:

\begin{equation}
    \frac{k_c}{d_a} \leq \frac{1}{4.10^7}
\end{equation}

\subsection{Parameter relations inferred from co-culture data}

The initial model described earlier is representative of the experimental settings of the co-culture conditions described in Kohda \textit{et al.}~\cite{kohda_vitro_2021}. At steady-state, the system \eqref{fullSystem} gives:

\begin{equation}\label{SEcoKohda}
[S_c]^* = 0 \quad \textrm{or} \quad [S_c]^* = K_{sc} \left ( 1 - \frac{d_{sc} [S_p]^*}{r_{sc} (C_1 + [S_p]^*)} \right)
\end{equation}
\begin{equation}\label{SAcoKohda}
[S_p]^* = 0 \quad \textrm{or} \quad [S_p]^* = K_{sp} \left ( 1 - \frac{d_{spb} [Amp_b]}{r_{sp} (C_{ab} + [Amp_b])} - \frac{d_{sph} [Amp_h]}{r_{sp} (C_{ah} + [Amp_h])} \right)
\end{equation}
\begin{equation}
{[Amp_b]}^* = \frac{k_c [S_c]^*}{d_a}
\end{equation}

Considering that what is observed experimentally after 48 hours of incubation is at steady-state, one can replace $[S_c]^*$ and $[S_p]^*$ with the experimental data point (\textit{S. epidermidis} $= 10^8$ CFU; \textit{S. aureus} $= 10^9$ CFU) in \eqref{SEcoKohda} and \eqref{SAcoKohda} to get the following parameter relation:

\begin{equation}\label{paramRKohda1}
\frac{d_{sc}}{r_{sc}} = \frac{3}{4.10^9} C_1 + \frac{3}{4}
\end{equation}
\begin{equation}\label{paramRKohda2}
\frac{2}{3} r_{sp} = \frac{d_{sph} [Amp_h]}{C_{ah} + [Amp_h]} + \frac{10^8 d_{spb} k_c}{d_a C_{ab}+10^8 k_c}
\end{equation}

By integrating the values found for $C_{ah}$ and $C_{ab}$, and the relations involving $d_{sph}$ and $d_{spb}$ into \eqref{paramRKohda2}, we end up with:

\begin{equation}\label{relation_da}
    d_a = 10^8 k_c \, \frac{56 + 31 [Amp_h]}{2.56\,(4-[Amp_h])} \quad \textrm{with } [Amp_h] < 4
\end{equation}

\section{Reduced model with 5 parameters} \label{sec:ReducedModel}

Using the previously mentioned experimental data, and assuming they represent steady state conditions of the initial model \eqref{fullSystem}, we have reduced the parametric dimension of the model from 13 to 5. Specifically, out of the original 13 parameters, we could define the values of 4 of them, and derive 4 functional dependencies from the values of the remaining parameters, as summarized in Table \ref{tabReducedModel}).

%\etg{This approach \ff{thus} leads us to consider correlations between parameters, guided by the data and the commonly used steady-state assumption on the experiments. To our knowledge, these correlations have not yet been observed experimentally and can also be considered as predictions \ff{of mathematical modeling.}}

\begin{table}
\caption{Summary of the parameter relations embedded in the reduced model.}\label{tabReducedModel}
\begin{center}
\begin{tabular}[H]{c|c}
\textbf{Parameter} & \textbf{Value or relation to other parameters}\\
\hline
$K_{sc}$ & $4.10^8$ \\
$K_{sp}$ & $3.10^9$ \\
$C_{ah}$ & 8 \\
$C_{ab}$ & 0.16 \\
$d_{sph}$ & $2 \, r_{sp}$\\
$d_{spb}$ & $\frac{5}{4} \, r_{sp}$\\
$d_{sc}$ & $\displaystyle{r_{sc} \left (\frac{3}{4.10^9} \; C_1 + \frac{3}{4} \right )}$ \\
$d_a$ & $\displaystyle{10^8 k_c \, \frac{56 + 31 [Amp_h]}{2.56\,(4-[Amp_h])}}$ with $[Amp_h] < 4$
\end{tabular}
\end{center}
\end{table}
 
 {In our \ff{reduced} skin microbiome model \ff{with AMPs} \eqref{fullSystem}, the parameters that remain unknown are thus:}
 \begin{itemize}
     \item 
 $r_{sc}$, the growth rate of $S_c$ which can reasonably take values between 0 and 2 $h^{-1}$ 
 following~\cite{czock_mechanism-based_2007,campion_pharmacodynamic_2005};
 \item
$r_{sp}$, the growth rate of $S_p$, taking similar values in the interval between 0 and 2 $h^{-1}$;
\item
$C_1$, the concentration of $S_p$ that induces half the maximum killing rate $d_{sc}$ (in $CFU.ASU^{-1}$)
and is thus bounded by the optimum concentration of $S_p$, i.e.~$K_{sp} = 3.10^9 \; CFU.ASU^{-1}$, as calculated in section \ref{subsec:Param_mono} from~\cite{kohda_vitro_2021};
\item
$k_c$, the production rate of $[Amp_b]$ chosen to take values between $0$ and $0.1 \; AU.h^{-1}.CFU^{-1}$, and shown to have a limited impact on the steady-state values in section \ref{subsec:Sensitivity};
\item$[Amp_h]$, the concentration in $AU.ASU^{-1}$ of AMPs produced by skin cells between 0 and 4 (equation \eqref{relation_da}).
 \end{itemize}

\etg{ This gives us the following reduced system:
 \begin{System}\label{reducedSystem}
\frac{d [S_c]}{dt} = r_{sc} [S_c] \left(  1 - \frac{[S_c]}{4.10^8}  - \frac{3}{4} \frac{(10^{-9} C_1 + 1 ) [S_p]}{C_1 + [S_p]} \right) \\ \\

\frac{d [S_p]}{dt} = r_{sp} [S_p] \left( 1 - \frac{[S_p]}{3.10^9}  - \frac{5 [Amp_b]}{0.64 + 4 [Amp_b]} - \frac{2 [Amp_h]}{8 + [Amp_h]} \right) \\ \\

\frac{d [Amp_b]}{dt} = k_c \left([S_c] - 10^8 [Amp_b] \frac{56 + 31 [Amp_h]}{2.56 (4 - [Amp_h])} \right)
\end{System}
 }
 
 \ff{A continuum of different parameter values exists however to fit either the experimental data of the pathogenic microbiome reported in~\cite{kohda_vitro_2021}, or what can be considered as a healthy microbiome phenotype. The landscape of those parameter values can nevertheless be explored using some formalization of the desired behaviors and global parameter optimization methods.}
 
 \subsection{\ff{Formal specification of dynamical behaviors in quantitative temporal logic}}\label{subsec:logic}

 \ff{
Quantitative temporal logics, such as Metric Temporal Logic MTL~\cite{AFH96jacm},
Signal Temporal Logic~\cite{MN04ftrtft}, or First-Order Linear Time Logic with real-valued linear constraints FO-LTL($\mathbb{R}_{lin}$)~\cite{RBFS11tcs,FR08tcs},
provide very expressive formal languages to specify  behaviors of continuous time dynamical systems~\cite{DM10formats}.

In particular, the definition of a continuous degree of satisfaction in the interval $[0,1]$ for FO-LTL($\mathbb{R}_{lin}$) formulae~\cite{RBFS09bi} 
is key to achieve several purposes:
first, to guide search in continuous optimization techniques to fit parameter values to satisfy model's temporal specification;
second, to evaluate the robustness of the model with respect to its specification and parameter variations; 
and third, to compute parameter sensitivity indices to determine the most influential parameters.

This is the method used here to fit our reduced model to the pathogenic phenotype reported in Kohda et al. experiments~\cite{kohda_vitro_2021},
and to a healthy phenotype.
We use the implementation of temporal logic FO-LTL($\mathbb{R}_{lin}$) in Biocham~\cite{RBFS11tcs} for this purpose and for sensitivity analyses\footnote{All computation results reported in this article have been obtained with \ff{a Biocham notebook} available in different formats at \url{https://lifeware.inria.fr/wiki/Main/Software\#TCS23}.}.
}

\ff{
For example, in the pathogenic experiments,
the property of interest is 
the reaching of an equilibrium around 48 hours of the bacterial populations 
with a predominance of the pathogenic bacteria of the following orders:
$[Sp]=10^9$ and $[Sc]=10^8$~\cite{kohda_vitro_2021}.
That first property is a simple curve fitting problem with two time points that can be formalized in FO-LTL($\mathbb{R}_{lin}$) by the following temporal logic formula with free variables $x1, y1, x2, y2$ for the population sizes 
\begin{equation}\label{foltlXY}
\textbf{F}(Time\sim 40 \wedge Sc = x1 \wedge Sp = y1 \wedge \textbf{F}(\textbf{G}(Sc = x2 \wedge Sp = y2)))
\end{equation}
and objective values 
\begin{equation}\label{objective1}
x1=10^8, y1=10^9, x2=10^8, y2=10^9
\end{equation}
(finally $\textbf{F}$) around time 40 
and (finally $\textbf{F}$ globally $\textbf{G}$) in all future time points 
(i.e.~at the last time point of a finite trace, 50 hours in the case of the experiments and of our model simulations). 

The temporal formula for fitting the model to a balanced microbiome expresses the dominance of the commensal bacteria over the pathogenic bacterial populations by at least one order of magnitude, as follows:
\begin{equation}\label{foltlrs}
\exists c~\exists p~ \textbf{F}(Time\sim 40 \wedge S_c > k*S_p \wedge S_c = c \wedge S_p = p \wedge \textbf{F}(\textbf{G}(r*S_p = p \wedge s*S_c = c)))
\end{equation}
with objective values 
\begin{equation}\label{objectivers}
    k=10,~r=1,~s=1.
\end{equation}

The abstraction by variables of the objective values makes it possible to compute 
first, the validity domain of the variables that satisfy the formula, as a finite union of polytopes~\cite{FS08tcs}, 
and second, a violation degree of the formula with respect to the objective values,
as the Haussdorf distance between the validity domain of the free variables and that objective point.
The landscape of parameter values satisfying the formula can then be explored in this setting by global parameter optimization. 

The continuous degree of satisfaction in $[0,1]$ of the specification 
is obtained as the inverse of the violation degree,
and can be used
as optimization criterion for parameter search~\cite{RBFS11tcs}.
In Biocham, the covariance matrix adaptive evolution strategy
(CMA-ES)~\cite{HO01ec} is used to search parameter values maximising the satisfaction degree of a FO-LTL formula given with respect to the objective values given for its free variables.

The satisfaction degree of a FO-LTL formula and objectives also provides
as computational measure of robustness of the model and sensitivity indices to the parameter values, by taking the mean satisfaction degree when varying the parameter values according to a normal distribution ${\cal N}(\mu, \sigma^2)$ around their nominal values $\mu$, with standard deviation $\sigma=COV.\mu$ given by a coefficient of variation $COV$.
}
 
\subsection{\ff{Model parameter fitting to pathogenic experiments and sensitivity analyses}}\label{subsec:Sensitivity}

In order to reproduce what was observed by Kohda et. al~\cite{kohda_vitro_2021}, 
that is \ff{the reaching of an equilibrium with} a dominant pathogenic population after \ff{40} hours, 
which can thus be considered as dysbiosis in our skin microbiome model, 
we \sout{can} fix \ff{the initial concentrations for $[S_c]$ and $[S_p]$ to the }
\sout{a relatively low concentration of Amp produced by the skin cells, e.g.~$Amph_h=1.5$. 
The} doses of \textit{S. epidermidis} and \textit{S. aureus} applied at the surface of the 3D epidermal equivalent at the beginning of the experiment (\ff{i.e.~}$10^5 ~ CFU/mL$ and $10^3 ~ CFU/mL$, respectively). \sout{ are used as the initial concentrations for $[S_c]$ and $[S_p]$, respectively}.

The parameterization problem consists in finding values for the remaining unknown parameters, $\ff{\Ah,~}r_{sc},~r_{sp},~C_1,~k_c$,
within the intervals of possible values described above,
\ff{and where $r_{sp}$ can be fixed to 1 arbitrarily.}
\sout{in order to reproduce the observations of the experiments~\cite{kohda_vitro_2021}.
The modeling platform Biocham~\cite{CFS06bi} we used for this study
%\footnote{All computation results reported in this article have been obtained with runnable Biocham files available in different formats at \url{https://lifeware.inria.fr/wiki/Main/Software\#TCS23}.},
provides original support for parameter search in high dimension, and robustness and sensitivity analyses, based on a formalization of the expected behaviour in quantitative temporal logic~\cite{RBFS11tcs}.
Here, the property of interest on a simulation trace is 
the stabilization of the bacterial population sizes to the values given by the experiments,
namely $[Sc]=10^8$ and $[Sp]=10^9$~\cite{kohda_vitro_2021},
at the time scale of the experiments around 48 hours.
These two values can be given as objectives for the free variables of the temporal logic formula
}
%begin{equation} %\label{foltlXY}
%F(Time\sim 40 \wedge Sc = x1 \wedge Sp = y1 \wedge F(G(Sc = x2 \wedge Sp = y2)))
%\end{equation}
\ff{\sout{which assigns the values of $Sc$ and $Sp$ finally (F) at time 40h, and finally globally (FG) at the end of the trace at the given time horizon of 50h.
}
}

\ff{Parameter optimization with respect to the property of reaching an equilibrium reached at time 40h (FO-LTL formula \ref{foltlXY}) with a dominance of the pathogenic bacterial population by one order of magnitude (objective values \ref{objective1})
shows the existence of a continuum of solutions.
Not surprisingly, the solutions found
assign small values to parameters $\Ah,~r_{sc}$ or $\C1$,
which could be anticipated from the structure of the model.
On the other hand, it is worth noting that no parameter values are found to reach the opposite equilibrium by one order of magnitude only, yet solutions are found for reaching a healthy microbiome with a dominance of the commensal bacteria over the pathogenic bacteria populations by 4 orders of magnitude\footnote{The Biocham notebook mentioned above shows the detailed results of parameter search and simulation figures on logscale.}.

In order to reproduce culture data, by opposition to in vivo measures, it is reasonable to fix parameter $\Ah$ to a relatively low value with respect to its range of typical values in the interval $[0,4]$.
When fixing $\Ah=1.5$ for instance,
parameter search shows the irrelevance of parameter $k_c$ 
and interestingly, a functional dependency of optimal parameter values $r_{sp}$
as a function of $C_1$.
Table \ref{C1} summarizes the optimal values found for $C_1$ and $r_{sp}$ in this setting.
}

\begin{table}
\caption{\ff{Optimal parameter values satisfying the reaching of an equilibrium after 40h (temporal formula \ref{foltlXY}) with a dominance of the pathogenic bacteria by one order of magnitude (objective values \ref{objective1}) when fixing $\Ah=1.5,~r_{sp}=1$. The parameter values retained are in bold.}}\label{C1}
\begin{center}
\begin{tabular}[H]{c|c|c|c|c|c|c|c|c}
$\C1$ & 1e3& 1e4 &1e5 &1e6 &\textbf{5e6} &1e7 &1e8 &1e9\\
$\rsc$& 1.3 &1 &0.76 &0.59 &\textbf{0.5} &0.46 &0.38 &0.34\\
\end{tabular}
\end{center}
\end{table}

\sout{
Among a continuum of solutions, due the low sensitivity of the property to the values of $\C1$ and $k_c$,}

\begin{figure}
    \centering
    \includegraphics[width=0.8\textwidth]{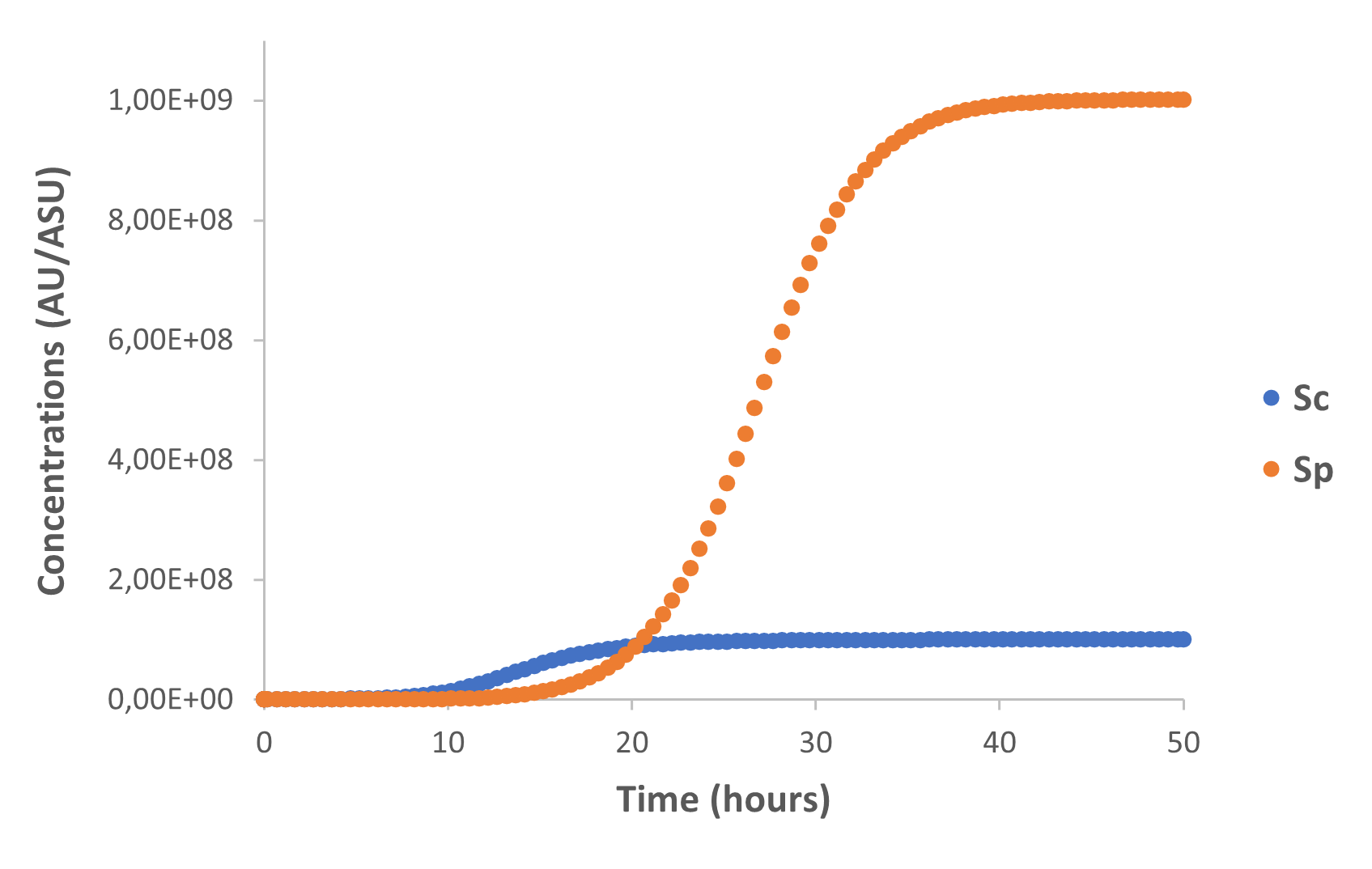}
    \caption{Numerical simulation of the reduced ODE model over 50 hours,
    with initial conditions $[S_c]=10^5,\ [S_p]=10^3,\ [Amp_b]=0$
    and parameter values $[Amp_h]=1.5,\ r_{sc} = 0.5, \, r_{sp} = 1 , \, C_1 = 5.10^6 , \, k_c = 0.01,$ to fit Kohda et al. co-culture data~\cite{kohda_vitro_2021} (Table \ref{tabExpDataKohda}).}
    \label{fig:Sim_Kohda}
\end{figure}

We choose \ff{accordingly} the parameter set $[\Ah=1.5, \,r_{sc} = 0.5, \, r_{sp} = 1 , \, C_1 = 5.10^6 , \, k_c = 0.01$
\ff{for fitting} those experiments.
Fig.~\ref{fig:Sim_Kohda} shows the result of a numerical simulation of our model 
with this parameter set which is in accordance to the co-culture experiments of Kohda et. al and does reproduce a consistent qualitative behavior~\cite{kohda_vitro_2021}.

The sensitivity analysis reported in Table \ref{tabSensitivityPathogenic} 
\ff{shows the expectation of the satisfaction degree of the temporal specification
when varying each parameter value one-at-a-time (OAT) using a normal distribution centered around its nominal value and some indicated coefficient of variation.
This} reveals that the dominance of the pathogenic population is not sensitive to the values of $\C1$ and $k_c$ which can vary by orders of magnitude, and that all model parameters can vary by 20\% without affecting the qualitative behaviour.
The highest sensitivity indices concern the growth rates ($r_{sp}$ and $r_{sc}$) 
which should remain with a COV below 20\%,
and the threshold concentration $C_{ah}$, 
followed by the optimum concentrations $K_{sc},~K_{sp}$ and $C_{ab}$.

\begin{table}
\caption{\ff{OAT and global s}ensitivity \ff{indices} of the pathogenic model to variations of all kinetic parameters and initial concentrations \ff{around their nominal values $\rsp=1, \rsc=0.5, \Ah=1.5, \C1=5e6, \kc=0.01, \Ksc=4e8, \Ksp=3e9, \Cab=0.16, \Cah=8$, for the equilibrium property of Kohda's co-culture experiments  (i.e.~formula \ref{foltlXY} with objective values \ref{objective1}). The most influential parameters given in decreasing order of OAT sensitivity indices with $COV=0.2$ are $\rsp,~\rsc,~\Cah,~\Ksc,~\Ksp,~\Cab,~\Ah,~\sc0,~\sp0,~\C1,~\kc$. Global sensitivity analysis provides an indication of the robustness of the model with respect to the equilibrium property for simultaneous variations of all model parameters around their nominal values with a COV of 10\%.}}\label{tabSensitivityPathogenic}
\begin{center}
\begin{tabular}[H]{c|c|c|c}
\textbf{Parameter} & \ff{\textbf{Nominal value}} & \ff{\textbf{COV}} & \textbf{Sensitivity index}\\
\hline
$\rsp$& $1$ &0.2&0.271\\
$\rsc$& $0.5$ &0.2&0.253\\
$\Cah$& $8$ &0.2&0.249\\
$\Ksc$& $4.10^8$ & 0.2& 0.194\\
$\Ksp$& $3.10^9$ & 0.2& 0.164\\
$\Cab$& $0.16$ &0.2&0.159\\
$\Ah$& $1.5$ &0.2&0.065\\
$\Ah$& $1.5$ &0.3&0.11\\
$\sc0$& $10^5$ &0.2&0.029\\
$\sc0$& $10^5$ &0.8&0.131\\
$\sp0$& $10^3$ &0.2& 0.026\\
$\sp0$& $10^3$ &0.5& 0.131\\
$\C1$& $5.10^6$ &0.2&0.021\\
$\C1$& $5.10^6$ &2&0.135\\
$\kc$& $$0.01 &0.2&0.01\\
$\kc$ &  $0.01$ &10 & 0.01\\
$\{\sp0, \sc0\}$ & \ff{$\{10^3, 10^5\}$}  &0.4 & 0.107\\
\ff{$\{\rsp, \rsc, \Ah, \C1, \kc\}$} & $"$& \ff{0.1} & \ff{0.206}\\
\ff{set of all 11 parameters} &$"$& \ff{0.1} & \ff{0.3}\\
\end{tabular}
\end{center}
\end{table}

\subsection{\ff{Parameter values fitting a balanced microbiome and sensitivity analyses}}

Our model can also be used to reproduce what is considered a balanced microbiome, corresponding to the commensal population being significantly more abundant than  the pathogenic one~\cite{kong_temporal_2012}. 
Given the high inter-individual variability, \ff{one cannot \sout{it is challenging to}} define precise target values for the population sizes corresponding to a balanced microbiome. Therefore, we will consider in the following analysis that the microbiome is balanced when \ff{it stabilizes with a} commensal population more abundant than the pathogenic one by at least one order of magnitude \ff{(see formula~\ref{foltlrs} with objectives~\ref{objectivers})}. 

\ff{Satisfying this property} requires modifying some parameter values to represent a less virulent pathogenic population, closer to the physiological context, given that the experiments from Kohda et. al~\cite{kohda_vitro_2021} were performed using a virulent methicillin-resistant \textit{S. aureus} strain.

\ff{Parameter search shows again a continuum of solutions for the parameter values, with optimal values for $\Ah$ around $2.5$, but no solution leading to a dominance of the commensals by just one order of magnitude, but by 4 orders of magnitude.}

\sout{For sensitivity analysis, we use a temporal logic formula which expresses that both bacterial populations stabilize at time 40h and 50h with a population size for $S_c$ greater than $S_p$ by one order of magnitude (independently of their absolute values).}
Sensitivity analysis \ff{reported in} Table \ref{tabSensitivity} reveals that the stable dominance of the commensal population is mostly sensitive to variations of 
$C_{ah}$, $C_{ab}$, $K_{sc}$ and interestingly $[Amp_h]$. \sout{with a coefficient of variation of 20\%}
On the other hand, the results are not sensitive to the initial bacterial population sizes
suggesting a form of perfect adaptation of the balance between both population.
\ff{In fact, this property follows from the absence of multiple non-degenerate steady states in this model, a property checked in the Biocham notebook of the model using the graphical requirements method for multistationarity of~\cite{BFS18jtb}. }

\begin{table}
\caption{\ff{OAT and global} sensitivity \ff{indices of the safe microbiome} model \sout{for safe microbiome (Fig.~\ref{fig:Sim_default})} to variations of the parameters and initial concentrations \ff{around their nominal values with a COV of 20\%} for the property of maintaining stable populations of bacteria at times 40h-50h with a predominance of the commensal bacteria by at least one order of magnitude \ff{(i.e.~formula \ref{foltlrs} with objective \ref{objectivers}). OAT sensitivity indices show the preponderant influence of parameters $\Cah,~\Cab,~\Ksc,~\Ah,~\rsc$ for maintaining a healthy balance, while the other kinetic parameters and the initial conditions appear much less important. Global sensitivity indices are an indication for the robustness of the model}.}\label{tabSensitivity}
\begin{center}
\begin{tabular}[H]{c|c|c|c}
\textbf{Parameters} & \ff{\textbf{Nominal value}} &\ff{\textbf{COV}} & \textbf{Sensitivity index}\\
\hline
$\Cah$& $8$ & 0.2 & 0.242\\
$\Cab$& $0.16$ & 0.2 & 0.207\\
$\Ksc$& $4.10^8$ & 0.2 & 0.188\\
$\Ah$& $2.5$ & 0.2 & 0.187\\
$\rsc$& $0.5$ & 0.2 & 0.007\\
$\Ksp$& $3.10^9$ & 0.2 & 0.005\\
$\sp0$& $10^3$ & 0.2 & 0.005\\
$\sc0$& $10^5$ & 0.2 & 0.005\\
$\rsp$& $0.5$ & 0.2 & 0.005\\
$\C1$& $2.10^9$ & 0.2 & 0.005\\
$\kc$& $0.01$ & 0.2 & 0.005\\
\ff{$(\rsp, \rsc, \Ah, \C1, \kc )$} &\ff{$"$}& 0.2& 0.184\\
\ff{tuple of all 11 parameters} &\ff{$"$}& 0.2 & 0.321\\
\end{tabular}
\end{center}
\end{table}

\begin{figure}
    \centering
    \includegraphics[width=0.8\textwidth]{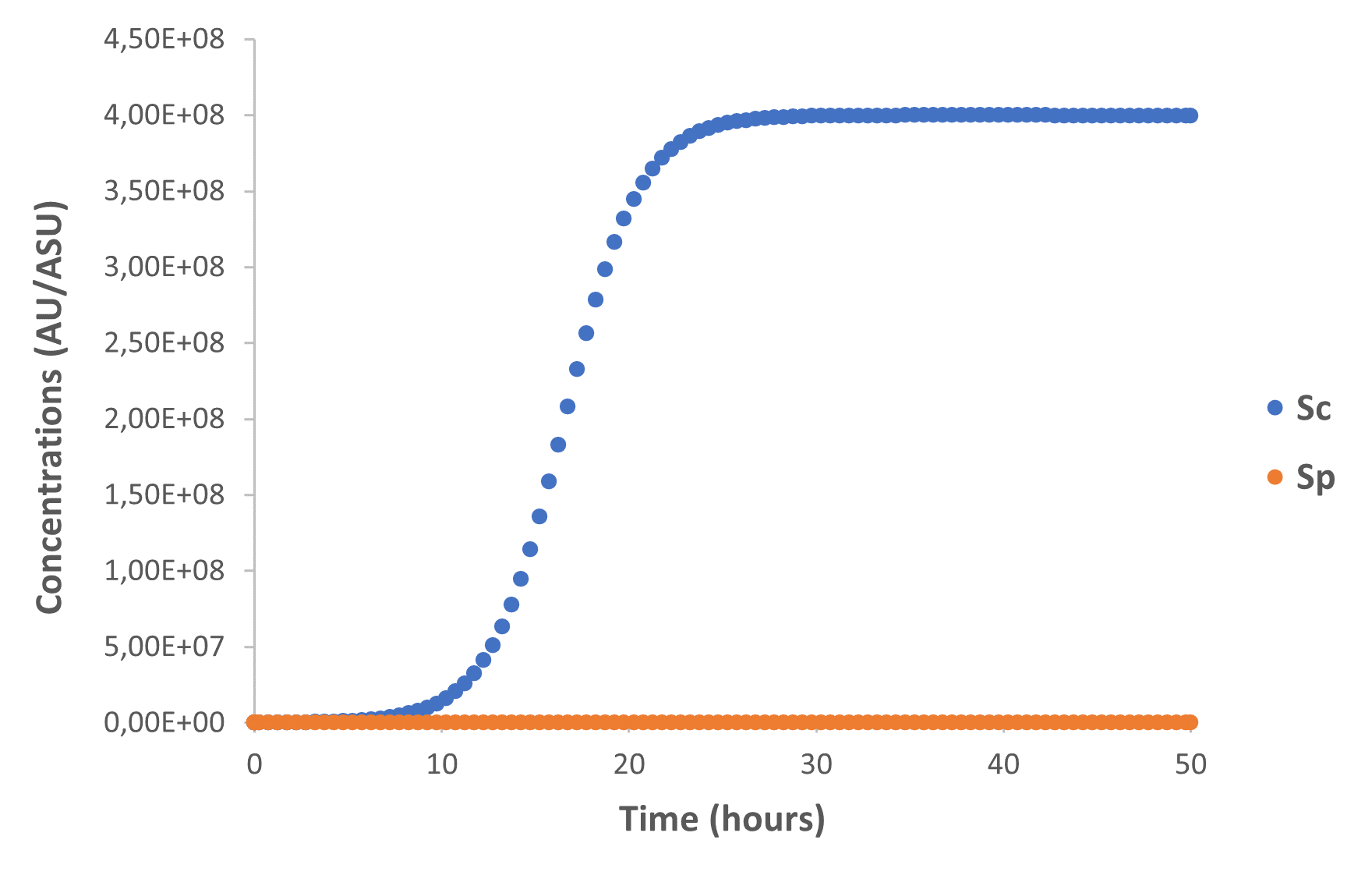}
    \caption{Numerical simulation of the reduced ODE model over 50 hours,
    with initial conditions $[S_c]=10^5,\ [S_p]=10^3,\ [Amp_b]=0$
    and parameter values $r_{sc} = r_{sp} = 0.5 , \, C_1 = 2.10^8 , \, k_c = 0.01 ,\ [Amp_h] = 3$ \ff{retained here for a healthy microbiome in a physiological context, with a value of $Sp=6.10^4$ at equilibrium, less than $Sc$ by 4 orders of magnitude.}}
    \label{fig:Sim_default}
\end{figure}

\ff{In order to get closer to physiological context however, we prefer to retain the slightly suboptimal parameter set} with
a higher production of AMPs by the skin cells, $[Amp_h] = 3$, to compensate for feedback loops or stimuli that might be missing in the 3D epidermal equivalent used,
and $r_{sp} = \rsc=0.5$, $C_1 = 2.10^8$.
Fig.~\ref{fig:Sim_default} shows a simulation trace obtained  under those conditions
which clearly \ff{satisfies the dominance property after 10h and reaching equilibrium around 25h, as shown also in the notebook with more details. \sout{indicates the dominance of the non-pathogenic population
under those conditions}}

\section{Conditions favoring the pathogenic population}

Whether the dysbiosis observed in AD is the cause or the result of the disease is unclear~\cite{kobayashi_dysbiosis_2015,koh_skin_2021}. Infants developing AD do not necessarily have more \textit{S. aureus} present on their skin prior to the onset of the disease compared to the healthy group~\cite{kennedy_skin_2017}. This suggests that atopic skin has some characteristics enabling the dominance of \textit{S. aureus} over the other species of the microbiome. To test this hypothesis, we investigate two changes of the skin properties observed in AD patients (skin surface pH elevation~\cite{eberlein-konig_skin_2000} and reduced production of AMPs~\cite{ong_endogenous_2002}), and their impact on the dominant species at steady-state. 

More specifically, we study the behavior of the system following the introduction of a pathogen and whether the pathogen will colonize the media depending on the initial concentrations of the bacterial populations and the particular skin properties mentioned before.

\subsection{Skin surface pH elevation}

According to Proksch~\cite{proksch_ph_2018}, the physiological range for skin surface pH is 4.1-5.8. However, in certain skin conditions, like AD, an elevation of this pH has been observed.
Dasgupta \textit{et al.}~studied \textit{in vitro} the influence of pH on the growth rates of \textit{S. aureus} and \textit{S. epidermidis}\cite{dasgupta_16502_2020}. Their experimental results show that, when the pH is increased from 5 to 6.5, the growth rate of \textit{S. epidermidis} is multiplied by 1.8, whereas the one of \textit{S. aureus} is multiplied by more than 4 (Table \ref{tabDasgupta}).

\begin{table}
\caption{Experimental data from Dasgupta et. al~\cite{dasgupta_16502_2020} showing the influence of pH on growth rates of \textit{S. epidermidis} and \textit{S. aureus}}\label{tabDasgupta}
\begin{center}
\begin{tabular}[H]{c|c|c|}
\multirow{2}{0.6cm}{\textbf{pH}} & \multicolumn{2}{c|}{\textbf{Growth rate ($\Delta$OD/hour)}}\\
\cline{2-3}
& \textit{\textbf{S. aureus}} & \textit{\textbf{S. epidermidis}}\\
\hline
5 & 0.03 & 0.05\\
5.5 & 0.04 & 0.07\\
6 & 0.09 & 0.08\\
6.5 & 0.13 & 0.09\\
7 & 0.14 & 0.10\\

\end{tabular}
\end{center}
\end{table}

Their data can be used to select values for the growth rates $r_{sc}$ and $r_{sp}$ in our model, corresponding to healthy skin with a skin surface pH of 5 and compromised skin with a pH of 6.5.
Because the experiments from Dasgupta \textit{et al.}~were performed \textit{in vitro} and the bacterial population sizes measured with optical density (OD) instead of CFU, the growth rates cannot be directly translated into $r_{sc}$ and $r_{sp}$.
\ff{Indeed, the experiments done by Dasgupta \textit{et al.}~were performed in vitro whereas the model is designed to represent the microbial dynamics in vivo. Even though in vitro experiments provide insightful results on the biological mechanisms studied, they also lack important factors such as interactions with the skin providing both nutrients and antimicrobial peptides, and skin renewal. In in vitro experiments the bacteria usually have access to an optimum amount of nutrients that can be renewed during the experiment or not. In vivo, the nutrients are provided by the skin. Depending on the skin state (dry, moisturized, with eczema,...) this quantity might not be optimal for bacterial growth. However, nutrients are continuously renewed due to new surface cells arising.
}

We use $r_{sc} = 0.5$ as the reference value for the commensal growth rate at pH 5, following on from previous simulation (Fig.~\ref{fig:Sim_default}). Maintaining the ratio between the two population growth rates at pH 5 and the multiplying factors following the pH elevation from Dasgupta \textit{et al.}~experimental data, we can define two sets of values for $r_{sc}$ and $r_{sp}$:

\begin{equation*}
\textrm{skin surface pH of 5} \quad \Rightarrow r_{sc} = 0.5, \, r_{sp} = 0.3
\end{equation*}
\begin{equation*}
\textrm{skin surface pH of 6.5} \quad \Rightarrow r_{sc} = 0.9, \, r_{sp} = 1.3
\end{equation*}

Considering the healthy skin scenario with a skin surface pH of 5, the influence of the bacterial populations initial concentrations on the dominant species after 50 hours is evaluate using the temporal logic formula:

$$F(\textrm{Time}==40 \wedge ([S_c] > u1 \, [S_p]) \wedge F(G([S_c] > u2 \, [S_p])))$$

where $u1$ and $u2$ are free variables representing the abundance factors between both populations,
evaluated at Time$= 40$ and at the last time point of the trace respectively (F stands for finally and G for globally at all future time points), i.e. at the time horizon of the experiments of 50 hours.
\ff{It is worth remarking that with that temporal formula, we do not impose that an equilibrium is reached since we want to focus primarily on the dominance criterion in this analysis.}

When given with an objective value, e.g. $u1=10$, the distance between that value
and the validity domain of the formula, i.e. the set of values for $u1$
that satisfy the formula, provides a violation degree which is used to evaluate 
the satisfaction degree of the property.

\begin{figure}
    \centering
    \includegraphics[width = 0.8\textwidth]{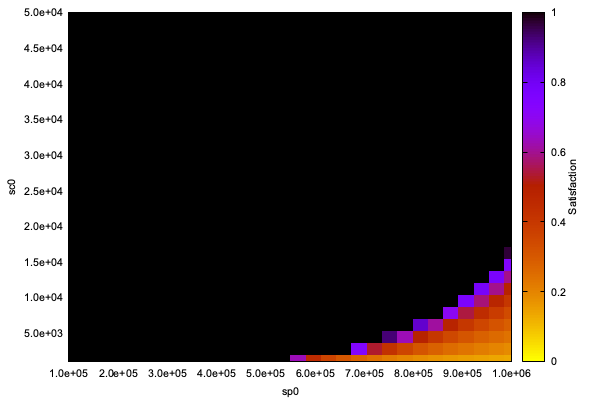}
    \caption{Landscape of satisfaction degree of the temporal formula corresponding to healthy skin with a skin surface pH of 5 ($r_{sc} = 0.5$ and $r_{sp} = 0.3$). The x and y axis represent variations of the initial quantities of $[S_p]$ and $[S_c]$, respectively. The color coding corresponds to the satisfaction degree of the temporal logic formula. Values used for the other parameters: $C_1 = 2.10^8$, $k_c = 0.01$, $[Amp_h] = 3$.}
    \label{fig:landscape_lowpH}
\end{figure}

Here, we evaluate how much the temporal formula $F(\textrm{Time}==40 \wedge ([S_c] > u1 \, [S_p]) \wedge F(G([S_c] > u2 \, [S_p])))$, $u1 \rightarrow 10, \, u2 \rightarrow 10$, is satisfied given variations of the initial concentrations of two populations (Fig.~\ref{fig:landscape_lowpH}).
The model predicts that, under the healthy skin condition, the commensal population will always dominate after 50 hours, except when introduced at a relatively low concentration ($<2.10^4$) while the initial concentration of the pathogenic population is high ($>5.10^5$).

\begin{figure}
    \centering
    \includegraphics[width = 0.8\textwidth]{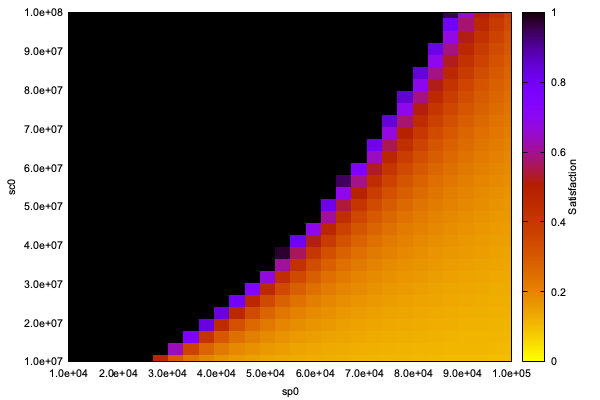}
    \caption{Landscape of satisfaction degree of the temporal formula corresponding to compromised skin with a skin surface pH of 6.5 ($r_{sc} = 0.9$ and $r_{sp} = 1.3$). The x and y axis represent variations of the initial quantities of $[S_p]$ and $[S_c]$, respectively. The color coding corresponds to the satisfaction degree of the temporal logic formula. Values used for the other parameters: $C_1 = 2.10^8$, $k_c = 0.01$, $[Amp_h] = 3$.}
    \label{fig:landscape_highpH}
\end{figure}

The model predicts a higher vulnerability of the skin regarding invading pathogens with an elevated skin surface pH. When evaluating the same temporal formula with growth rates values corresponding to a skin surface pH of 6.5, we observe that even when the initial concentration of commensal is high ($>10^7$), the pathogenic population is able to colonize the skin when introduced at a concentration as low as $3.10^4$ (Fig.~\ref{fig:landscape_highpH}).
Such predictions highlight the protective effect of the skin surface acidic pH against the invasion of pathogenic bacteria. 

\ff{It was expected that an elevated pH would favor the pathogens. It is worth remarking however that the magnitude of the pathogens dominance compared to the small difference in growth rates ($\rsp$ 44\% higher than $\rsc$) was surprising to us. Modeling more explicitly the skin surface pH would require a much more complex model since the acidic skin surface pH depends on the concentration in Natural Moisturizing Factor (NMF). NMF is constituted of small molecules resulting from the degradation of other molecules such as filaggrin deeper within the skin. The exact mechanisms involved in establishing and maintaining the acidic skin surface pH are still not completely understood. 
Short of modeling the pH impact on bacteria on a molecular scale, for which there is not much data available, changes in pH can only be modeled through a change in bacterial growth rates, death rates or carrying capacities ($\Ksp$ and $\Ksc$). Here, we chose to impact the growth rates based on the data that was available.}

\subsection{Reduced production of skin AMPs}

\begin{figure}
    \centering
    \includegraphics[width = 0.8\textwidth]{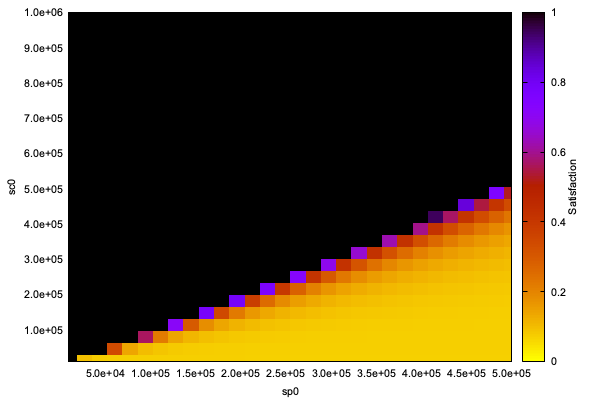}
    \includegraphics[width = 0.8\textwidth]{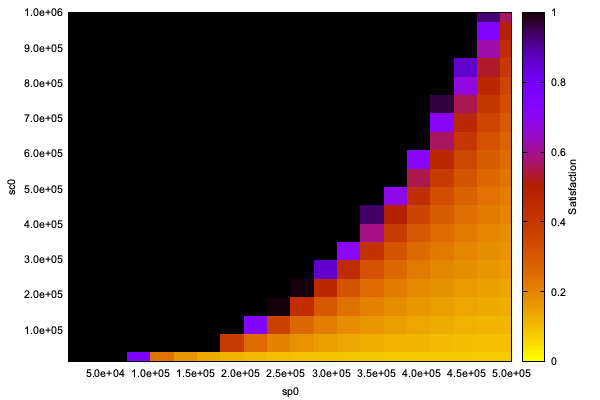}
    \caption{Landscape of satisfaction degree of the healthy condition formula with a low concentration of human AMPs on the upper graph ($[Amp_h] = 0.5$) and a high concentration at the bottom ($[Amp_h] = 3$). The x and y axis represent variations of the initial quantities of $[S_p]$ and $[S_c]$ respectively. The color coding corresponds to the satisfaction degree of the temporal logic formula. Values used for the other parameters: $r_{sc} = r_{sp} = 0.5$, $C_1 = 2.10^8$, $k_c = 0.01$.}
    \label{fig:landscape_Amph}
    
\end{figure}

As mentioned before, human keratinocytes constitutively produce AMPs as a defense against pathogens. In atopic dermatitis, the expression of AMPs is dysregulated, leading to lower concentration levels of AMPs in the epidermis~\cite{nakatsuji_antimicrobials_2017}. 
Similarly to the analysis done for skin surface pH, our model can be used to study how the skin microbiome reacts to modulation of the AMPs production by the skin cells.
Two situations are considered: an impaired production of AMPs by the skin cells ($[Amp_h] = 0.5$) and a higher concentration with $[Amp_h] = 3$.
Using the same methodology as in the case of skin surface pH, the temporal logic formula $F(\textrm{Time}==40 \wedge ([S_c] > u1 \, [S_p]) \wedge F(G([S_c] > u2 \, [S_p])))$, $u1 \rightarrow 10, \, u2 \rightarrow 10$, is evaluated for variations of the initial concentrations of both populations for $[Amp_h] = 0.5$ and $[Amp_h] = 3$ (Fig.~\ref{fig:landscape_Amph}).\\

The model predicts a slightly protective effect of $Amp_h$ regarding the colonization of the skin by a pathogenic population, for low initial concentrations. However when both populations are introduced in high concentrations, the increase of $[Amp_h]$ appears to have the opposite effect of facilitating the colonization by the pathogenic population.

This mitigated effect \etg{is unexpected from a biological standpoint and} might be due to the presence of $[Amp_h]$ in the constraint related to the degradation rate of $[Amp_b]$ (Equation \eqref{relation_da})\etg{. \ff{That equation was derived from a} steady-state assumption made on our model to relate it to some available experimental observations. It resulted in making $[Amp_b]$ degradation rate $d_a$ proportional to the concentration in $[Amp_h]$. There is no evidence in the literature that an increased concentration in human AMPs might enhance the degradation of bacterial AMPs. The predicted non-linear protective effect of human AMPs on the skin should \ff{thus} be further investigated, \ff{either experimentally or by refinement of our} model. \ff{In case of discrepancy with new experimental evidence, that model prediction could serve as a guide} towards a modification of the model structure or a reconsideration of the previous \ff{mathematical} steady-state assumption made on the experimental data. \sout{and deserves further investigation.}}

\section{Quasi-stability revealed by simulation on a long time scale}

\begin{figure}
    \centering
    \includegraphics[width = 0.8\textwidth]{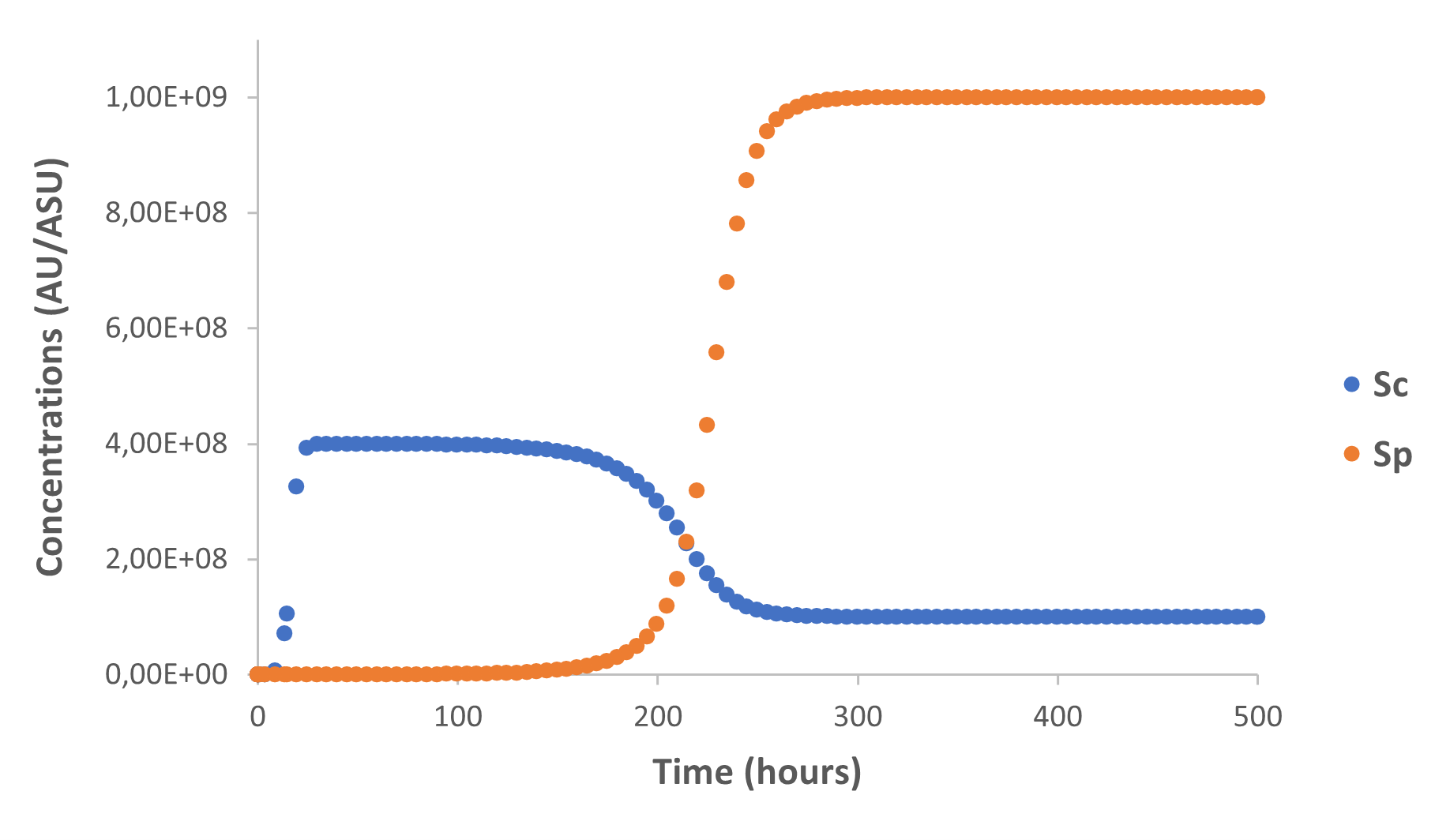}
    \caption{{Numerical simulation of the reduced ODE model on a longer time scale of 500 hours, with the same initial concentrations and parameter values as in Fig.~\ref{fig:Sim_default}, showing an inversion of the dominant bacterial population after 220 hours.}}
    \label{fig:metastab}
\end{figure}

Interestingly, by extending the simulation time horizon to a longer time scale of 500 hours,
one can observe a quasi-stability phenomenon, shown in Fig.~\ref{fig:metastab}.
The \ff{equilibrium} observed in Fig.~\ref{fig:Sim_default} at the relevant time scale of 50 hours of the experiments, is thus not a \ff{mathematically stable} state, 
but a quasi-stable state, 
that slowly evolves, with $\displaystyle{\frac{d[S_c]}{dt} \neq 0}$ and $\displaystyle{\frac{d[S_p]}{dt} \neq 0}$,
towards a true stable state of the model reached after 300 hours
in which the population density are reversed.

The $S_c$ population almost reaches its optimum capacity $K_{sc}$ after approximately 30 hours and stays relatively stable for around 100 hours more, that is over 4 days, which can reasonably be considered stable on the microbiological time scale. Meanwhile, the $S_p$ population is kept at a low concentration compared to $S_c$, even though it is continuously increasing and eventually leading to its overtake of $S_c$.

By varying the parameters values, it appears that this quasi-stability phenomenon emerges above a threshold value of $2.5$ for $[Amp_h]$, that is for almost half of its possible values (see Section \ref{sec:ReducedModel}). 

Such phenomena of quasi-stability, also called meta-stability when the switches result from stochasticity~\cite{bovier1970metastability} \ff{\sout{are standard notions of} appear in different forms in the \ovr{theory of deterministic} dynamical systems ~\cite{TK08neuron,RSNGW15cmsb,samal2016geometric,samal2018metastable,desoeuvres_CMSB2022,morozov2020long}.
\sout{They are particularly well-studied in the case of oscillatory systems,% for which analytical solutions exist, 
e.g.~in models of brain activity~\cite{TK08neuron}.
They are also central in computational systems biology for simplifying biochemical reaction networks. For instance, they provide a model reduction method based on the identification of different regimes 
corresponding to different preponderant terms of the ODEs, for which simplified dynamics can be defined, and chained within a hybrid automaton~\cite{RSNGW15cmsb}.}}
The observation of this phenomenon in our \ff{simple population dynamics model is however quite surprising and} deserves an original mathematical analysis to understand its sources.

\begin{figure}
    \centering
    \includegraphics[width = 0.8\textwidth]{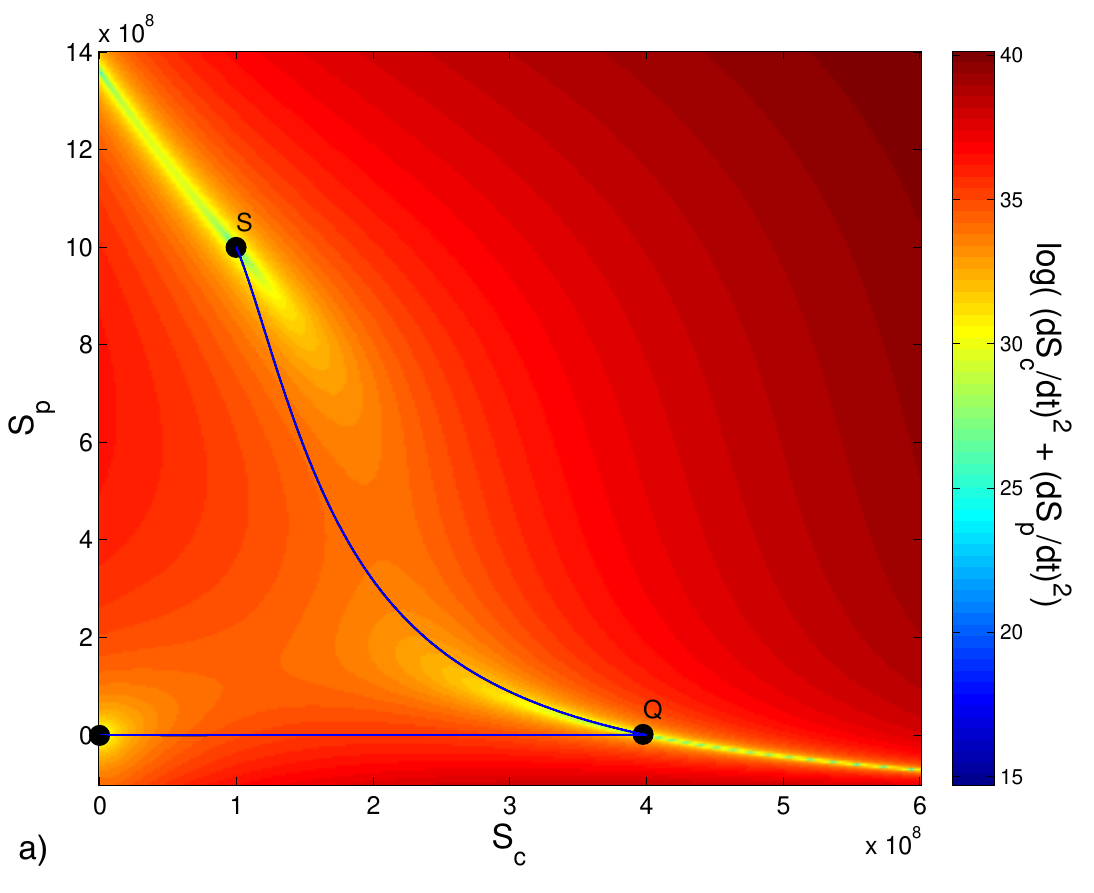}
    
    \includegraphics[width = 0.8\textwidth]{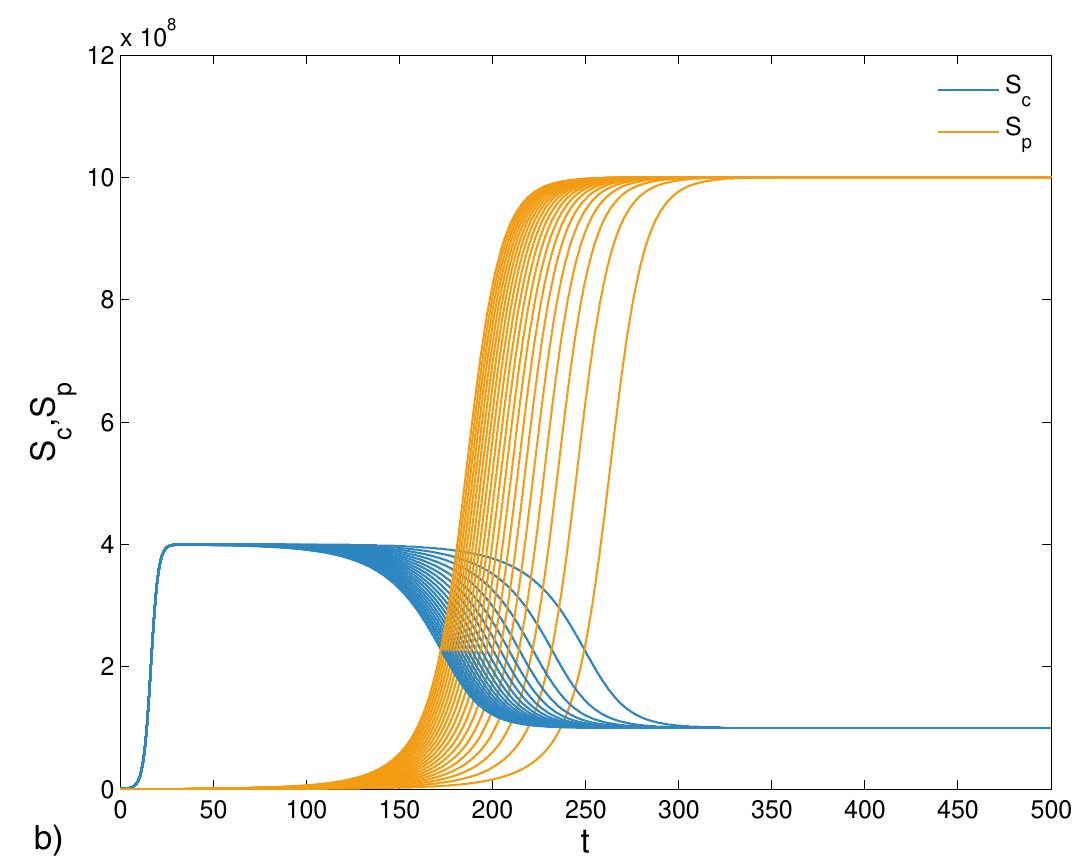}
    
    \caption{Quasi-stability versus stability illustrated for the 2D skin microbiome model obtained by eliminating the fast variable $[Amp_b]$
    and using the parameters of Fig.~\ref{fig:Sim_default}.
    a) A typical orbit leaves the neighborhood of the unstable state ($S_c=0$,$S_p=0$), reaches the quasi-stable state Q where it stays for a long time
    and ends with the stable state S. b) The corresponding time series with different values of $S_p(0)$ shows that for a long time $S_p$ is constant whereas 
    $S_c$ increases exponentially but very slowly starting with very low values. The transition to the stable steady state occurs when
    $S_c = 2 \times 10^8$ and the time spent in the quasi-stable state is given by $\log( 2 \times 10^8 / S_c(0) )/\alpha$ where
    $S_c(0)$ and $\alpha$ are the initial values and effective exponential growth rate of the pathogen, respectively. }
    \label{fig:phase_portrait}
\end{figure}

\section{Mathematical Analysis of Quasi-stability versus Stability in Population Dynamics}\label{sec:math}

\subsection{Quasi-stability versus stability}

Dynamical systems theory is widely applied in biology. Notions such as attractors and  steady
states are systematically used for the analysis of biological models. Stable steady states (point attractors) 
are zero velocity states attracting all neighboring orbits (state S in Figure~\ref{fig:phase_portrait}~a)). 
\ff{Quasi-stability phenomena are particularly well-studied in the case of oscillatory systems,% for which analytical solutions exist, 
e.g.~in models of brain activity~\cite{TK08neuron}.
They are also central in computational systems biology for simplifying biochemical reaction networks. For instance, they provide a model reduction method based on the identification of different regimes 
corresponding to different preponderant terms of the ODEs, for which simplified dynamics can be defined
\cite{samal2015geometric,rvg,kruff2021algorithmic,desoeuvres2022reduction},
and chained within a hybrid automaton~\cite{RSNGW15cmsb,desoeuvres_CMSB2022}.}
\ff{\sout{Less known are the quasi-stable states
that we tentatively} 
\ovr{The simplified dynamics takes place on attractive, normally
hyperbolic invariant manifolds, sometimes defined as }
 quasi-steady states like state Q in Figure~\ref{fig:phase_portrait}~a. \ovr{These states attract neighboring orbits} along some fast, stable 
directions, and evolve very slowly along other directions
\cite{kruff2021algorithmic,desoeuvres2022reduction}. 
}

Quasi-stable states were discussed
for models in ecology~\cite{morozov2020long}, but are certainly more general. 
A non-exhaustive list of nonlinear phenomena 
leading to quasi-stability include:
i) slow attractive, normally hyperbolic invariant manifolds, that exist for slow/fast systems~\cite{wiggins1994normally,kruff2021algorithmic,desoeuvres2022reduction};
ii) saddle points with slow unstable directions, where slowness can be due to the proximity  of a saddle-node 
bifurcation~\cite{morozov2020long}; and
iii) ghost attractors, also due to the proximity of a saddle-node bifurcation~\cite{morozov2020long}.

\subsection{Quasi-stability of the reduced parametric model}

We first look for slow invariant manifolds by using tropical geometry scaling methods introduced elsewhere \ovr{\cite{noel2012tropical,noelgvr,soliman2014constraint,radulescu2020tropical}}.
By these methods we identify fast and slow variables and use quasi-steady state as zero order approximations
to invariant manifolds~\cite{samal2015geometric,rvg,kruff2021algorithmic,desoeuvres2022reduction}. 

\ovr{For the sake of simplicity we rename the model variables and parameters
from \eqref{fullSystem}. The analysis of the model is performed using 
formal parameters. However, we use the nominal parameter values from  Table~\ref{tabReducedModel} and Figure 3 to compute parameter orders
of magnitude that are considered fixed throughout this
section. These choices lead to
}
%From \eqref{fullSystem} and \ovr{using the Table~\ref{tabReducedModel}}, and after renaming variables and parameters, 
%e.g.~constant $[Amp_h]$ is now denoted $k_c$, 
%we obtain
\begin{eqnarray}
\dot x_1 &=&  k_1 x_1 \left( 1  - \frac{x_1}{k_2} - \frac{k_3 x_2 }{  k_4 + x_2} 
 \right)  , \notag \\
\dot x_2 &=&    k_5 x_2 \left(k_9  -  \frac{x_2}{k_6} - \frac{ k_7 x_3 }{ k_8 + x_3} \right),\notag \\
\dot x_3  &=&   k_{10} (   x_1  -   k_{11} x_3   ),  \label{eq:renamed}
\end{eqnarray}
\ovr{where $x_1 = [S_c], x_2 = [S_p], x_3 = [Amp_b]$, $k_1=r_{sc}=0.5$,
$k_2=K_{sc}=4 \times 10^8$, 
$k_3=d_{sc}/r_{sc}=9/10$, 
$k_4=C_1=2\times 10^8$, 
$k_5=r_{sp}=0.5$,$k_6=K_{sp}=3\times 10^9$, 
$k_7=d_{spb}/r_{sp}=5/4$,
$k_8=C_{ab}=0.16$,  
$k_9=1- \displaystyle{\frac{d_{sph}[Amp_h]}{r_{sp}(C_{ah}+[Amp_h])} = 5/11}$, 
$k_{10}=K_c=0.01$, 
$k_{11}=d_a/K_c=\displaystyle{\frac{149}{256} \times 10^{10}}$.}

The first step of tropical scaling is to rescale parameters as powers of a small positive parameter $\epsilon$, 
$$k_i = \bar k_i \epsilon^{\gamma_i},$$
where $\gamma_i = round(\log(k_i)/\log(\epsilon))$ 
\cite{samal2015geometric,rvg,kruff2021algorithmic,desoeuvres2022reduction}.

For instance, for $\epsilon = 1/11$, we have 
$\bar k_1 = 0.5, \gamma_1 = 0$,
$\bar k_2 = 1.86, \gamma_2 = -8$,
$\bar k_3 = 9/10, \gamma_3 = 0$,
$\bar k_4 = 0.933, \gamma_4 = -8$,
$\bar k_5 = 0.5, \gamma_5 = 0$,
$\bar k_6 = 1.27, \gamma_6 = -9$,
$\bar k_7 = 5/4, \gamma_7 = 0$,
$\bar k_8 = 1.76, \gamma_8 = 1$,
$\bar k_9 = 5/11, \gamma_9 = 0$,
$\bar k_{10} = 1.21, \gamma_{10} = 2$,
$\bar k_{11} = 2.46, \gamma_{11} = -9$.
Here the parameters $k_1$, $k_5$, $k_4$, $k_{10}$ and $k_{11}$ were chosen as in  Fig.~\ref{fig:Sim_default}
to illustrate quasi-stability.

By doing so, parameters are represented by their orders of magnitude $\gamma_i$. \ovr{In the generic situation when the rescaled, zero order  parameters $\bar k_i$ are independent, 
$\gamma_i$ determine alone the orders of 
magnitude of the system's timescales
\cite{samal2015geometric,rvg,kruff2021algorithmic,desoeuvres2022reduction}. However, when $\bar k_i$ are dependent (for
instance when particular expressions of these parameters are small in absolute value) the
orders of magnitude of the timescales may also depend on $\bar k_i$. We will see that our problem corresponds to
the non-generic case.} 

The second step of the tropical scaling method is to introduce all of order variables 
$x_i = y_i \epsilon^{a_i}$ and to consider that the orders of magnitude $a_i$ of the variables satisfy
tropical equilibration conditions. 

\ovr{Although the tropical scaling  method has been
most often presented in the framework of polynomial ODEs~\cite{samal2015geometric,rvg,kruff2021algorithmic,desoeuvres2022reduction}, early tropicalization ideas use 
rational ODEs~\cite{noel2012tropical}. The idea is to decompose the 
r.h.s. of each ODE into a sum of rational terms that have constant signs.
For each ODE, a tropical equilibration condition means that its r.h.s. contains at least two such rational } terms of opposite signs and dominant (smallest) order of magnitude. 
\ovr{The orders of magnitude of rational terms are computed using the} tropical algebra (with addition replaced by minimum and 
multiplication by plus). \ovr{Therefore,  the tropical
equilibration} conditions read:

\begin{eqnarray}
0 &=& \min (-\min(-a_2-8,0),a_1 +8) \notag \\
0 &=& \min(a_2 + 9, -\min(1-a_3,0)) \notag \\
0 &=& a_1 - a_3 + 9 
\end{eqnarray}
Using the last equation we obtain the equivalent system
\begin{eqnarray}
0 &=& \min (-\min(-a_2-8,0),a_1 +8) \notag \\
0 &=& \min(a_2 + 9, -\min(-a_1-8,0)) \notag \\
a_3 &=& a_1  + 9 \label{tropical_system}
\end{eqnarray}
The system \eqref{tropical_system} has
 two one dimensional branches of solutions
\begin{description}
    \item[I.] $a_1 = -8$, $a_2 > -8$, $a_3 = 1$.
    \item[II.] $a_2 = -9$, $a_1 > -8$, $a_3 =a_1  + 9$.
\end{description}
The first branch corresponds to large $x_1$ (commensal bacteria) and small $x_2$ (pathogens), whereas
the second branch corresponds to small $x_1$ and large $x_2$. 

For the initial conditions in this study the branch I is appropriate. The rescaled system is then
\begin{eqnarray}
\dot y_1 &=&  \bar k_1 y_1 \left( 1  - \frac{y_1}{\bar k_2}  - \frac{\bar k_3 y_2 \epsilon^{a_2+8} }{ \bar k_4  + y_2 \epsilon^{a_2+8}} 
\right)  , \notag \\
\dot y_2 &=&   \bar k_5 y_2 \left(\bar k_9  -\epsilon^{a_2+9}  \frac{\bar{y_2}}{\bar k_6} - \frac{\bar k_7 y_3 }{\bar k_8 + y_3} \right),\notag \\
\dot y_3  &=& \epsilon^{-7} \bar k_{10} (   y_1  -  \bar k_{11} y_3   ).  \label{eq:slowfast}
\end{eqnarray}

From the rescaled system one can see that $y_3$ is fast and $y_1$ and $y_2$ are slow.
%As a matter of fact, 
%\ovr{a similar calculation shows that} the same property
%is valid for the branch II as well. 
Therefore, one can reduce the system of 3 ODEs to a system of 2 ODEs by using the 
quasi-steady approximation for the variable $y_3$, namely $y_3 = y_1 / \bar k_{11}$. This reduction is robust as it is valid
for low and high values of the sub-populations. 

On the branch I, because $a_2 + 8 > 0$ the terms scaling like $\epsilon^{a_2+8}$ and $\epsilon^{a_2+9}$ can be neglected. 
The 2D reduced order model reads
\begin{eqnarray}
\dot y_1 &=&  \bar k_1 y_1 \left( 1  
- \frac{y_1}{\bar k_2}  \right)  , \notag \\
\dot y_2 &=&   \bar k_5 y_2 \left(\bar k_9 - \frac{\bar k_7 y_1 }{\bar k_8 \bar k_{11} + y_1} \right). \label{eq:reduced}
\end{eqnarray}
In the system \eqref{eq:reduced} the first equation is decoupled from the
second. In a short time, $y_1 \to \bar k_2$ and then $y_2$ follows the
equation
\begin{equation}
\dot y_2 =   \bar k_5 y_2 \left(\bar k_9 - \frac{\bar k_7  \bar k_2}{\bar k_8 \bar k_{11}  + \bar k_2} \right) 
\label{eq:exponential}
\end{equation}
The condition for this to happen is 
that $y_1$ is faster than $y_2$ which reads
\begin{equation}
0 <
\bar k_5 (\bar k_9 - \frac{\bar k_7  \bar k_2}{\bar k_8 \bar k_{11}  + \bar k_2} ) << \bar k_1
\label{eq:condition}
\end{equation}
Using the numerical values of the parameters
we find
$$\bar k_5(\bar k_9 - \frac{\bar k_7  \bar k_2}{\bar k_8 \bar k_{11}  + \bar k_2}) = 0.0790 \bar k_1.$$ 
The variable $y_2$ is slow with a characteristic time 
$$\left[ \bar k_5  \left(\bar k_9 - \frac{\bar k_7  \bar k_2}{\bar k_8 \bar k_{11}  + \bar k_2} \right)\right]^{-1} = (0.5 \times 0.0790)^{-1} = 25.32 h ,$$ meaning that $x_2$ (like $y_2$) needs $58.32 h$ to change by a decade. That explains the observed quasi-stability. 

\ovr{The numerical value of the characteristic time was obtained from the nominal parameters used in Figure~3. For different values of the zero order rescaled parameters $\bar k_i$ one should verify our condition \eqref{eq:condition} in order to conclude about 
quasi-stability. Furthermore, if orders of magnitude of the parameters change, the tropical scaling and thus the entire
analysis change. A full formal parametric analysis, discussing all possible
tropical scalings depending on parameter orders of magnitude, is possible but left for a different publication.  
}

\subsection{Non-genericity of this form of quasi-stability}

In contrast to slow-fast systems, where quasi-stability is generic (depends \ovr{only on} orders of magnitudes of 
parameters) here one has non-generic slowing down because of compensation between zero order parameters
$\bar k_i$ (condition \eqref{eq:condition}). 

The effective dynamics in the quasi-stable state consists in very slow exponential growth of the pathogen
population $x_2$ with the effective 
rate 
\begin{equation}
\alpha = \bar k_5(\bar k_9 - \frac{\bar k_7  \bar k_2}{\bar k_8 \bar k_{11}  + \bar k_2}),
\end{equation}
 while the commensal population $x_1$ is constant, see also Figure~\ref{fig:phase_portrait}.
 % $x_1 \approx k_2$.
%The state Q is thus approximately defined by $x_1 = k_2, 0 < x_2 < \epsilon^{-8}$.
When the pathogen population becomes too large the conditions needed for the existence of the 
branch I are no longer satisfied
and all the above approximations break down. The system exits quasi-stability and evolves towards the
stable steady state S (Figure~\ref{fig:phase_portrait}). 
As a matter of fact, the time needed to exit $Q$ corresponds to the time needed for $x_2$ to reach $k_4$.
When this happens, the term  $\frac{\bar k_3 y_2 \epsilon^{a_2+8} }{ \bar k_4  + y_2 \epsilon^{a_2+8}}$  can no longer be 
neglected in the first equation of the system \eqref{eq:reduced}. This time depends on the initial value
$x_2(0)$, scales like $[\log(k_4)  -\log(x_2(0))]\alpha^{-1}$, and becomes infinite for $x_2(0)=0$.

\subsection{General conditions in population dynamics models}
One would expect that this slowing down mechanism by compensation of zero order parameters occurs more 
generally in models of antagonistic species. In order to illustrate this, let us consider the following
model describing competition of $N$ species:
\begin{equation}\label{general}
\dot x_i = r_i x_i ( 1 - \frac{x_i}{K_i} - \sum_{j\neq i} \frac{\beta_{ij} x_j}{\theta_{ij} + x_j} ), 1 \leq i \leq N,
\end{equation}
where $x_i$ are the species concentrations. 

\ovr{We consider that  the orders of magnitude of $r_i$, $\alpha_i$, $K_i$, $\beta_{ij}$ and $\theta_{ij}$ are all zero. More generally, for parameters with different orders of magnitude,
we should first use tropical scaling to obtain a rough 
fast/slow decomposition of the ODE system. 
In the general case, \eqref{general} represents the slowest dynamics resulting from tropical rescaling and describing interactions
among slow species, direct or
mediated by fast species. }

Let us further consider two categories of species 
defined by indices subsets $S_1$ and $S_2$, where 
$S_1 \cap S_2 = \emptyset$, $S_1 \cup S_2 = \{1,\ldots,N\}$. 
$S_1$ includes  the indices of saturated species, whose concentrations will tend to the
capacities $K_i$ and $S_2$ includes 
the indices of species whose initial concentrations are close to zero, initially. 
For simplicity, we  consider that species $S_1$ do not compete one another, i.e. $\beta_{ij}=0$ for
$i,j \in S_1$. Under these conditions, the dominating terms in the ODE model are
\begin{eqnarray}
\dot x_i &=& r_i x_i ( 1 - \frac{x_i}{K_i} ), \text{for } i \in S_1, \label{eqS2} \\
\dot x_i &=& r_i x_i ( 1  - \sum_{j \in S_1} \frac{\beta_{ij} x_j}{\theta_{ij} + x_j} ), \text{for } i \in S_2. \label{eqS1} 
\end{eqnarray}
The approximate Eq.\eqref{eqS2} was obtained by neglecting the terms 
$\sum_{j \in S_2} \frac{\beta_{ij} x_j}{\theta_{ij} + x_j}$, which is valid as long as $x_j << \theta_{ij}$ 
for all $i\in S_1$, $j\in S_2$.
 
According to \eqref{eqS2} the species $S_1$ stabilize to saturation levels $K_i$ and according to 
\eqref{eqS1} the species $S_2$ increase exponentially  with rates
$r_i ( 1  - \sum_{j \in S_1} \frac{\beta_{ij} K_j}{\theta_{ij} + K_j} ) $. Therefore,  the quasi-stability conditions read:
\ovr{\begin{eqnarray}
&0 < r_i(1 - \sum_{j \in S_1} \frac{\beta_{ij}  K_j}{\theta_{ij} + K_j})  << 1, \text{ for all } i \in S_2 \label{qs_cond1} \\
&x_j(0) << \theta_{ij}, \,
\text{for all } i\in S_1, j\in S_2
\label{qs_cond2}
\end{eqnarray}

\ovr{The non-negativity and the much-less-than conditions in \eqref{qs_cond1} ensure that concentrations of species from $S_2$ increase, and that the increase is slow, respectively. The much-less-than condition in \eqref{qs_cond2} ensures
that the approximate equation \eqref{eqS2} is initially valid.  }
If \eqref{qs_cond1},\eqref{qs_cond2} }are satisfied, then the saturated species reach almost constant levels $x_i = K_i,\, i\in S_1$  after a short transient, and the 
non-saturated species increase very slowly. 
The lifetime of this quasi-stable state is the smallest time for some
non-saturated species $i$ to reach the threshold level $\theta_{ij}$ in which case \eqref{eqS2} is no longer true.   This lifetime
reads
$$
\min_{i \in S_2, j\in S_1} 
%\left\{ 
\frac{\log(\theta_{ij})-\log(x_i(0)) }{
r_i( 1 - \sum_{l \in S_1} \frac{\beta_{il}  K_l}{\theta_{il} + K_l} )}.
%\right\}.
$$

\section{Conclusion}

In the perspective of identifying conditions which might favor or inhibit the emergence of pathogenic populations in the skin microbiome,
we have developed a simple ODE model of skin microbiome with 3 variables and 13 parameters,
reduced to 5 parameters by using, on the one hand, published data from the literature,
and on the other hand, steady state reasoning for taking into account available data from biological experiments.
Our bacterial population model is generic in the sense that we did not 
take into account the peculiarities of some specific bacterial populations,
but on some general formulas of adversary population dynamics and influence factors.
We showed through sensitivity analyses that our model predictions are reasonably  robust
with respect to parameter variations. \etg{\sout{
Such model-based analyses can thus lead to insights about potential treatment strategies aiming at restoring a dysbiotic condition.}

The presented model predicts that a) elevation of skin surface pH, creates favorable conditions for the emergence and colonization of the skin by the opportunistic pathogen populations to the detriment of commensals and b) the concentration of human AMPs is non-linearly affecting the balance between pathogens and commensals. The first prediction is in line for example with current knowledge of the effect of pH on staphylococcus aureus in AD. The model can be used to generalize this observation to other pathogens and can justify the use of optimal pH levels in topical formulations. The second prediction is novel and the specific non-linearity can be used to design strategies that optimize AMP levels on the skin surface using appropriately formulated topical treatments.}

Perhaps surprisingly, we also showed that this simple model exhibits a quasi-stability phenomenon over a large range of biologically relevant parameter values, revealed by allowing the simulation  to continue for times one order of magnitude longer than the reported time of the biological experiments.
This observation questions the existence and importance of quasi-stability phenomena 
in real biological processes, whereas a natural assumption made in mathematical modeling, 
and model fitting to data, is that the experimental data are observed in states corresponding to real stable states of the mathematical model.

Using tropical scaling, we have shown that, in contrast to well-studied slow-fast systems, quasi-stability in our model
does not depend on the orders of magnitude of the parameters,
but results from a non-generic slowing-down of zero order parameters\ff{, for which our analysis provides analytic estimates of the timescale dependence on the parameters}.
\ff{Finally, those conditions for quasi-stability, obtained by tropical algebra analysis of our skin microbiome model,}  \ff{have been shown to generalize} to population dynamics models, with an arbitrary number of species, for their potential applications in other domains.

\subsubsection*{Acknowledgments.} We are grateful to Mathieu Hemery, Aurélien Naldi and Sylvain Soliman for interesting discussions on this work, \ff{and to the anonymous reviewers for their useful comments to improve the presentation of our results}.

\bibliographystyle{splncs04}
\bibliography{TFRSS23tcs}

\end{document}